\documentclass[a4paper,11pt]{article}
\pdfoutput=1
\usepackage{jheppub}
\usepackage{color,bm,comment,braket}
\usepackage[T1]{fontenc}
\newcommand{\del}{\partial}
\newcommand{\beq}{\begin{eqnarray}}
\newcommand{\eeq}{\end{eqnarray}}
\newcommand{\dis}{\displaystyle}
\newcommand\+{\dagger}

\newcommand{\tr}{\mathop{\mathrm{tr}}}
\newcommand{\SU}{\text{SU}}
\newcommand{\U}{\text{U}}
\newcommand{\rmi}{\text{i}}

\title{Topological term, QCD anomaly, and the $\eta^{\prime}$ chiral soliton lattice in rotating baryonic matter}

\author{Kentaro Nishimura and Naoki Yamamoto}

\affiliation{Department of Physics, Keio University, \\
3-14-1 Hiyoshi, Yokohama, Japan}

\emailAdd{nishiken.a6@keio.jp}
\emailAdd{nyama@rk.phys.keio.ac.jp}

\abstract{We study the ground states of low-density hadronic matter and high-density color-flavor locked color superconducting phase in three-flavor QCD at finite baryon chemical potential under rotation. We find that, in both cases under sufficiently fast rotation, the combination of the rotation-induced topological term for the $\eta^{\prime}$ meson and the QCD anomaly leads to an inhomogeneous condensate of the $\eta^{\prime}$ meson, known as the chiral soliton lattice (CSL). We find that, when baryon chemical potential is much larger than isospin chemical potential, the critical angular velocity for the realization of the $\eta^{\prime}$ CSL is much smaller than that for the $\pi_0$ CSL found previously. We also argue that the $\eta^{\prime}$ CSL states in flavor-symmetric QCD at low density and high density should be continuously connected, extending the quark-hadron continuity conjecture in the presence of the rotation.}

\begin{document}
\maketitle
\flushbottom

\section{Introduction}
Investigating the phase diagram of quantum chromodynamics (QCD) at finite temperature $T$ and/or baryon chemical potential $\mu_{\rm B}$ is one of the important problems in the Standard Model of particle physics. In recent years, QCD matter not only at finite $T$ and $\mu_{\rm B}$, but also under rotation ${\bm \Omega}$ has attracted much attention. Experimentally, it has been reported that quark-gluon plasmas produced in noncentral heavy ion collision experiments at Relativistic Heavy Ion collider (RHIC) have the largest vorticity observed so far, of order $10^{22}/ {\rm s}$ \cite{STAR:2017ckg}. There have also been theoretical studies on the phases of QCD matter under rotation mostly using the effective models of QCD \cite{Chen:2015hfc,Ebihara:2016fwa,Jiang:2016wvv,Chernodub:2016kxh,Chernodub:2017ref,Liu:2017zhl,Zhang:2018ome,Wang:2018zrn,Chen:2019tcp}.

Recently, it has been shown in ref.~\cite{Huang:2017pqe}, based on a systematic low-energy effective theory, that the ground state of two-flavor QCD at finite $\mu_{\rm B}$ and isospin chemical potential $\mu_{\rm I}$ under sufficiently fast rotation is an inhomogeneous condensate of the $\pi_0$ meson, called the chiral soliton lattice (CSL). Generally, the CSL is a periodic array of topological solitons that spontaneously breaks parity and translational symmetries. This CSL is a universal state of matter in that it appears in various systems from condensed matter physics to high-energy physics, such as chiral magnets \cite{Dzyaloshinsky:1964dz,togawa2012chiral}, cholesteric liquid crystals \cite{Gennes}, and QCD at finite $\mu_{\rm B}$ in an external magnetic field \cite{Brauner:2016pko}; see also refs.~\cite{Brauner:2019rjg,Brauner:2019aid} for the CSL in QCD-like theories without the fermion sign problem.

As summarized in table~\ref{tab:csl}, the realization of the CSL typically requires three essential ingredients: a Nambu-Goldstone (NG) boson field (say $\theta$) associated with some symmetry $G$, a total derivative term for $\theta$, and explicit breaking of the symmetry $G$ that provides a mass term for $\theta$. In chiral magnets, the so-called Dzyaloshinskii-Moriya (DM) interaction gives a total derivative term for the magnon field $\varphi$ of the form ${\bm D} \cdot {\bm \nabla} \varphi$, where ${\bm D}$ is the DM vector \cite{kishine2015theory}, and an external magnetic field gives a mass term for $\varphi$ due to the explicit breaking of the spin rotational symmetry. In QCD matter, the Wess-Zumino-Witten (WZW) type terms in a magnetic field \cite{Son:2004tq, Son:2007ny} or under rotation \cite{Huang:2017pqe} lead to total derivative terms for $\pi_0$, and the quark mass provides a mass term for $\pi_0$ due to the explicit breaking of chiral symmetry. Note that the total derivative term for $\pi_0$ under rotation is present only at \emph{both} finite $\mu_{\rm B}$ and $\mu_{\rm I}$ \cite{Huang:2017pqe} (see also section~\ref{subsec:cve-topo} below), and as a result, the CSL of the $\pi_0$ meson is not realized in baryonic matter under rotation.

\begin{table}[tbp]
\begin{tabular}{|c|c|c|c|} \hline
Physical system & NG boson  & Total derivative term & Explicit symmetry breaking \\ \hline \hline
\begin{tabular}{c}
Chiral magnet 
\end{tabular}
& Magnon $\varphi$ & DM interaction ${\bm D} \cdot {\bm \nabla} \varphi$ & Magnetic field 
\\ \hline 
\begin{tabular}{c}
QCD (${\bm B}, \mu_{\rm B}$)
\end{tabular}
& $\pi_0$ & WZW-type term $\mu_{\rm B} {\bm B} \cdot {\bm \nabla} \pi_0$ & Quark masses  \\ \hline 
\begin{tabular}{c}
QCD (${\bm \Omega}, \mu_{\rm B}, \mu_{\rm I}$)
\end{tabular}
& $\pi_0$ & WZW-type term $\mu_{\rm B} \mu_{\rm I} {\bm \Omega} \cdot {\bm \nabla} \pi_0$ & Quark masses  \\  \hline 
\begin{tabular}{c}
QCD (${\bm \Omega}, \mu_{\rm B}$)
\end{tabular}
& $\eta^{\prime}$ & WZW-type term $\mu_{\rm B}^2 {\bm \Omega} \cdot {\bm \nabla} \eta'$ & QCD anomaly and quark masses \\ \hline 
\end{tabular}
\caption{Examples of the CSL in chiral magnets and QCD matter.} \label{tab:csl}
\end{table}

In this paper, we show that, in three-flavor QCD at finite $\mu_{\rm B}$ under sufficiently fast rotation, another CSL-type ground state is realized---the CSL of the $\eta'$ meson, or simply the $\eta'$ CSL. The ingredients for the existence of the $\eta'$ CSL are a new WZW-type topological term for $\eta'$ at finite $\mu_{\rm B}$ under rotation (see eqs.~(\ref{eta-cve-topological}) and (\ref{cve-topo-CFL})) and the QCD anomaly (or the ${\rm U}(1)_{\rm A}$ anomaly), which provides an additional mass term for $\eta'$. In both cases of low-density hadronic matter and high-density color-flavor locked (CFL) color superconducting phase, we analytically derive the critical angular velocities above which the $\eta'$ CSL states are realized and the CSL-type ground-state configurations. In particular, we find that, in the regime $\mu_{\rm B} \gg \mu_{\rm I}$, the critical angular velocity for the realization of the $\eta^{\prime}$ CSL is much smaller than that for the $\pi_0$ CSL found in ref.~\cite{Huang:2017pqe}. We also argue that these $\eta^{\prime}$ CSL states at low density and high density should be continuously connected in flavor-symmetric QCD, extending the quark-hadron continuity conjecture \cite{Schafer:1998ef, Hatsuda:2006ps,Yamamoto:2007ah} in the presence of the rotation.

This paper is organized as follows. In sections \ref{sec:ChPT} and \ref{sec:csl}, we construct the low-energy effective theories for low-density hadronic matter and high-density CFL phase under rotation, respectively, and show that their ground states under sufficiently fast rotation are the $\eta^{\prime}$ CSL states. Section~\ref{sec:discussion} is devoted to discussion and conclusion.

Throughout this paper, we consider QCD at finite $\mu_{\rm B}$ and at zero temperature.
We set the angular velocity along the $z$ direction, ${\bm \Omega} \equiv \Omega \hat{\bm z}$, without loss of generality. The effect of this rotation can be expressed by the metric
\begin{gather}
{\rm d}s^2 = (1 - \Omega^2 r_{\perp}^2){\rm d}t^2 - 2g_{0i}{\rm d}t {\rm d}x^i 
\label{metric-rot} \,,
\end{gather}
where $r_{\perp} \equiv \sqrt{x^2 + y^2}$ is the distance from the $z$ axis and $g_{0i}$ satisfies $\Omega = \epsilon^{z jk} \del_jg_{0k}$, or explicitly,
\begin{gather}
g_{\mu \nu} = \left(
\begin{array}{cccc}
1-\Omega^2 (x^2 + y^2) & \Omega y & \dis -\Omega x & 0 \\
\Omega y & -1 & 0 & 0 \\
-\Omega x & 0 & -1 & 0 \\
0 & 0 & 0 & -1
\end{array}
\right) \,.
\end{gather}
We also assume that $r_{\perp} \Omega < 1$ so that the velocity of the boundary does not exceed the speed of light.

\section{Low-density hadronic matter under rotation}
\label{sec:ChPT}
In this section, we consider the low-energy effective theory---the chiral perturbation theory (ChPT)---for low-density hadronic matter under rotation. 

We start from massless three-flavor QCD and we ignore the ${\rm U}(1)_{\rm A}$ anomaly for a moment. (We will consider the effects of quark masses and the ${\rm U}(1)_{\rm A}$ anomaly later.) In this case, QCD has the $\U(3)_{\rm R} \times \U(3)_{\rm L}$ chiral symmetry:
\begin{gather}
\label{chiral-rotation-quark} 
q_{\rm R} \to {\rm e}^{-\rmi \lambda_0\theta_0^{\rm R}}V_{\rm R}q_{\rm R}\,, \qquad  
q_{\rm L} \to {\rm e}^{-\rmi \lambda_0\theta_0^{\rm L}}V_{\rm L}q_{\rm L} \,, 
\end{gather}
where $V_{\rm R,L}\equiv \exp(-\rmi \lambda_A\theta_A^{\rm R, L})$ are the $\SU(3)_{\rm R,L}$ transformations for right- and left-handed quarks $q_{\rm R, L}$ and 
$\lambda_a$ are the $\U(3)$ generator with the normalization $\tr(\lambda_a\lambda_b)=2\delta_{ab}$. Here and below, the indices $A$ and $a,b,c$ stand for $A=1,2, \cdots, 8$ and $a,b,c=0,1,\cdots , 8$, respectively.
We assume that this chiral symmetry is spontaneously broken down to the vector $\U(3)_{\rm V}$ symmetry in the vacuum and low-density hadronic matter. As a result, the nonet of NG bosons appears. We can parametrize the field of the nonet mesons by the $\U(3)$ matrix $U$,
\begin{gather}
U = \Sigma \, \exp \left(\frac{\rmi \lambda_0\eta^{\prime}}{f_{\eta^{\prime}}}\right)\,, \qquad 
\Sigma = \exp \left(\frac{{\rm i}\lambda_A\pi_A}{f_{\pi}}\right) \,, 
\end{gather}
where $f_{\pi, \eta^{\prime}}$ are the decay constants of the octet and singlet mesons. 
The field $\Sigma$ transforms under $\SU(3)_{\rm R} \times \SU(3)_{\rm L}$ as
\begin{gather}
\label{U-transformation}
\Sigma \to V_{\rm L}\Sigma V_{\rm R}^{\+}\,, 
\end{gather}
while $\eta^{\prime}$ transforms under $\U(1)_{\rm A}$ as
\begin{gather}
\label{eta'-transformation}
\eta^{\prime} \to \eta^{\prime} + 2f_{\eta^{\prime}}\theta_0 \,,
\end{gather}
where $\theta_0 \equiv \theta_0^{\rm R} = -\theta_0^{\rm L}$. 

The kinetic terms invariant under eqs.~(\ref{U-transformation}) and (\ref{eta'-transformation}) up to ${\mathcal O}(\del^2)$ are written as
\begin{gather}
{\mathcal L}_{\rm kin}^{\rm ChPT}
=\frac{f_{\pi}^2}{4} g^{\mu \nu} {\rm tr}(\del_{\mu}\Sigma \del_{\nu}\Sigma^{\dag}) + \frac{1}{2}g^{\mu \nu}\del_{\mu}\eta^{\prime}\del_{\nu}\eta^{\prime} \,, \label{kin}
\end{gather}
where $g^{\mu \nu}$ is an inverse matrix of the metric $g_{\mu \nu}$ in eq.~(\ref{metric-rot}).
This effective theory is based on the expansion in terms of the small parameter $p/(4 \pi f_{\pi, \eta'}) \ll 1$ with $p$ being the momentum. In the following, we will also take $\Omega/(4 \pi f_{\pi, \eta'})$ as a small expansion parameter and we will consider the leading-order contributions of $\Omega$ in the effective theory.

\subsection{Chiral vortical effect and topological term} 
\label{subsec:cve-topo}
We review the topological terms in QCD matter under the global rotation derived in ref.~\cite{Huang:2017pqe}.%
\footnote{See also ref.~\cite{Manes:2019fyw} for a related recent work.}
The idea is based on the matching of the anomaly-induced transport phenomenon called the chiral vortical effect (CVE) \cite{Vilenkin:1979ui,Vilenkin:1980zv,Son:2009tf,Landsteiner:2011tf,Landsteiner:2012kd,Landsteiner:2016led}, which is a current along the direction of a vorticity or rotation in relativistic matter, between the microscopic theory (QCD) and low-energy effective theory (ChPT).

We consider QCD with finite chemical potentials $\mu_a$ ($a=0,1,\cdots ,8$) associated with conserved charges ${\bar q}\gamma^0\lambda_a q$. Under the global rotation ${\bm \Omega}$, the system exhibits the axial vector currents $\bm{j}^5_a$ in the direction of the rotation axis:
\begin{align}
{\bm j}^5_a
= N_{\rm c}\frac{d_{abc}}{2\pi^2} \mu_b\mu_c {\bm \Omega} \,, \qquad d_{abc} \equiv \frac{1}{2} \tr[\lambda_a \{ \lambda_b, \lambda_c \} ] \,,
\label{cve}
\end{align}
where $N_{\rm c}$ is the number of colors and the transport coefficient $d_{abc}$ is the same as the chiral anomaly coefficient \cite{Landsteiner:2011tf,Landsteiner:2012kd,Landsteiner:2016led}. Because this expression of the CVE is exact independently of the energy scale similarly to the chiral anomaly, it must also be reproduced in terms of the NG bosons in the ChPT.

To derive the effective Lagrangian that reproduces the CVE, we consider a local axial rotation,
\begin{gather}
q \to {\rm e}^{-{\rm i}\lambda_a\theta_a \gamma_5}q, 
\end{gather}
where the parameter $\theta_a$ depends on $x^{\mu}$. Under the infinitesimal transformation, the action of QCD changes as
\begin{gather}
\delta S_{\rm QCD}
= \int {\rm d}^4x \, \del_{\mu}\theta_a \cdot j^{5 \mu}_a \,,
\end{gather}
where we used $\del_{\mu}j^{5 \mu}_a = 0$ in the chiral limit. In terms of the NG bosons, the chiral rotation is expressed by
\begin{gather}
\label{chiral-rotation-NG}
\pi_A \to \pi_A + 2f_{\pi} \theta_A\,, \qquad \eta^{\prime} \to \eta^{\prime} + 2f_{\eta^{\prime}}\theta_0 \,.
\end{gather}
Thus, the change of the effective action under the transformation is
\begin{gather}
\delta S_{\rm EFT}
= \int {\rm d}^4x \, \left[\del_{\mu} \left(\frac{\delta \pi_A}{2f_{\pi}} \right) \cdot j^{5 \mu}_A 
+ \del_{\mu}\left(\frac{\delta \eta^{\prime}}{2f_{\eta^{\prime}}}\right)\cdot j^{5\mu}_{a=0}
\right] \,,
\end{gather}
where we define $\delta \pi_A \equiv 2f_{\pi} \theta_A$ and $\delta \eta^{\prime}\equiv 2f_{\eta^{\prime}}\theta_0$. The exactness of the CVE leads to the matching condition $\delta S_{\rm QCD} = \delta S_{\rm EFT}$, from which we arrive at the topological term \cite{Huang:2017pqe}, 
\begin{gather}
\label{anom_cve}
{\mathcal L}_{\rm CVE}
= N_{\rm c} \frac{d_{Abc}}{4\pi^2f_{\pi}} \mu_b\mu_c {\bm \nabla}\pi_A \cdot {\bm \Omega} 
+ N_{\rm c}\frac{d_{0bc}}{4\pi^2f_{\eta^{\prime}}} \mu_b\mu_c {\bm \nabla}\eta^{\prime} \cdot {\bm \Omega} \,.
\end{gather}

This derivation is similar to the anomaly matching condition used to derive the WZW term in the ChPT \cite{Wess:1971yu,Witten:1983tw}, which is responsible for, e.g., the $\pi_0 \rightarrow 2 \gamma$ decay. Moreover, in the external magnetic field at finite $\mu_{\rm B}$, there is an additional WZW-type term of the form, $\sim \mu_{\rm B} {\bm B} \cdot {\bm \nabla} \pi_0$ \cite{Son:2004tq,Son:2007ny}, which is responsible for the chiral magnetic effect \cite{Vilenkin:1980fu, Nielsen:1983rb, Fukushima:2008xe}.

In the case of QCD at finite $\mu_{\rm B}$, the only nonvanishing component of eq.~(\ref{anom_cve}) is
\begin{gather}
\label{eta-cve-topological}
{\mathcal L}_{\rm CVE}^{\eta^{\prime}}
= \frac{\mu^2_{\rm B}}{4\pi^2f_{\eta^{\prime}}} \sqrt{\frac{2}{3}} {\bm \nabla}\eta^{\prime} \cdot {\bm \Omega} \,,
\end{gather}
where we used $\sqrt{\frac{2}{3}}\mu_0 = \frac{1}{N_{\rm c}} \mu_{\rm B}$ and $N_{\rm c} = 3$.
On the other hand, the topological term for the $\pi_0$ meson appears only at finite $\mu_{\rm B}$ {\it and} $\mu_{\rm I}$ as \cite{Huang:2017pqe}
\begin{gather}
\label{pi-cve-topological}
{\mathcal L}_{\rm CVE}^{\pi_0}
= \frac{\mu_{\rm B} \mu_{\rm I}}{2\pi^2f_{\pi}} {\bm \nabla}\pi_0 \cdot {\bm \Omega} \,.
\end{gather} 
Therefore, the topological term (\ref{eta-cve-topological}) is dominant compared with eq.~(\ref{pi-cve-topological}) in the regime $\mu_{\rm B} \gg |\mu_{\rm I}|$.

\subsection{$\U(1)_{\rm A}$ anomaly and quark mass terms}
\label{sec:ChPT-anom}
We next include the effects of explicit chiral symmetry breaking in the ChPT: finite quark masses and the ${\rm U}(1)_{\rm A}$ anomaly.

We first consider the corrections in the ChPT due to the quark masses. We introduce the quark mass term in the original QCD Lagrangian,
\begin{gather}
{\mathcal L}_{\rm mass}^{\rm QCD}
= - {\bar q}_{\rm L}Mq_{\rm R} - {\bar q}_{\rm R}M^{\dag}q_{\rm L} \,, \label{quark mass} \\
M = {\rm diag}(m_{\rm u}, m_{\rm d}, m_{\rm s}) \label{mass matrix} \,.
\end{gather}
In order to determine the form of the corresponding effective Lagrangian at low energy, we promote the quark mass matrix $M$ into a spurion field and 
require that $M$ transforms under the chiral rotation (\ref{chiral-rotation-quark}) as
\begin{gather}
M \to {\rm e}^{-2\rmi \lambda_0\theta_0}V_{\rm L}MV_{\rm R}^{\dag} \label{transformation-mass-matrix} \,,
\end{gather}
such that eq.~(\ref{quark mass}) is invariant under this transformation. The effective Lagrangian involving $M$ and $U$ that is invariant under the extended symmetry in eqs.~(\ref{U-transformation}), (\ref{eta'-transformation}) and (\ref{transformation-mass-matrix}) up to ${\mathcal O}(M)$ reads
\begin{gather}
{\mathcal L}_{\rm mass}^{\rm ChPT}
=B \tr(MU + {\rm h.c.}) \label{mass-ChPT} \,,
\end{gather}
where the parameter $B$ is not determined by the symmetry alone. 

It is also known that the ${\rm U}(1)_{\rm A}$ part of the chiral symmetry $\U(3)_{\rm R} \times \U(3)_{\rm L}$ is explicitly broken by the QCD anomaly. This results in the symmetry breaking ${\rm U}(1)_{\rm A} \to {\mathbb Z}_{6}$ in three-flavor QCD. Let us incorporate this ${\rm U}(1)_{\rm A}$ anomaly into the ChPT. Such an anomalous term breaks ${\rm U}(3)_{\rm R} \times {\rm U}(3)_{\rm L}$ symmetry, but preserves ${\rm SU}(3)_{\rm R} \times {\rm SU}(3)_{\rm L} \times {\rm U}(1)_{\rm B} \times {\mathbb Z}_{6}$. Then, one of the options is (see, e.g., \cite{Aoki:2014moa})
\begin{gather}
{\mathcal L}_{\rm anom}
= \frac{a}{2} \left({\rm det}U + {\rm det}U^{\+}\right)
\label{ChPT-det-anomaly} \,,
\end{gather}
where $a$ represents the strength of the ${\rm U}(1)_{\rm A}$ anomaly. 
As $\det \Sigma=1$, eq.~(\ref{ChPT-det-anomaly}) is
\begin{gather}
{\mathcal L}_{\rm anom}
= a \, {\rm cos}\left(3\sqrt{\frac{2}{3}}\frac{\eta^{\prime}}{f_{\eta^{\prime}}} \right) \label{QCDanom-ChPT}
\end{gather}
which gives the additional mass term of the $\eta^{\prime}$ meson as
\begin{gather}
\delta m^2_{\eta^{\prime}}=\frac{6a}{f^2_{\eta^{\prime}}} \,.
\end{gather}

An important remark is in order here. It is known that the Lagrangian~(\ref{ChPT-det-anomaly}) does not satisfy the proper $1/N_{\rm c}$ counting rules in the large-$N_{\rm c}$ expansion, e.g., for the quartic $\eta'$ interaction \cite{Witten:1980sp}. One could instead write down the Lagrangian with the proper counting rules as
\begin{gather}
{\mathcal L}_{\rm anom}' = -\frac{f^2_{\pi}a}{4N_{\rm c}} \left({\rm i} \ln \det U \right)^2
\label{axial_anomaly} \,,
\end{gather}
where $f_{\pi} = {\mathcal O}(N^{1/2}_{\rm c})$ and $a$ is some constant with $a={\mathcal O}(N^0_{\rm c})$, and the normalization of the prefactor is chosen following ref.~\cite{Witten:1980sp}. 
However, it turns out that eq.~(\ref{axial_anomaly}) explicitly breaks spatial translational symmetry due to the cusp singularities in the potential (see appendix \ref{app:Eta-largeN} for details), and so it does not work out for our purpose to study its spontaneous breaking by the CSL state. For this reason, we will adopt the Lagrangian~(\ref{ChPT-det-anomaly}) below. In fact, the detailed form of the Lagrangian for the ${\rm U}(1)_{\rm A}$ anomaly will be irrelevant to our discussions, but the only essential piece will be just the strength of the ${\rm U}(1)_{\rm A}$ anomaly---the coefficient $a$ in the case of the Lagrangian (\ref{ChPT-det-anomaly}). In passing, we also note that mathematically the same form of the Lagrangian as eq.~(\ref{ChPT-det-anomaly}) appears in CFL phase at high density, as we will see in section~\ref{subsec:EFT-CFL}.

\subsection{Ground state in the chiral limit}
\label{subsec:GSofChPT}
In the following, we will concentrate on the $\eta^{\prime}$ meson and set $U={\rm e}^{{\rm i}\phi}$ with $\phi \equiv \sqrt{6}\eta^{\prime}/(3f_{\eta^{\prime}})$. 
For convenience, we use the cylindrical coordinate $(r,\theta,z)$. 

We first consider the case where the ${\rm U}(1)_{\rm A}$ anomaly is sufficiently large so that the quark mass term is negligible. (We will consider the effects of the quark mass in section~\ref{subsec:GS(finite quark masses)}). In this case, adding eqs.~(\ref{kin}), (\ref{mass-ChPT}), (\ref{ChPT-det-anomaly}) and (\ref{anom_cve}) together, we obtain the effective Hamiltonian in the static limit as
\begin{align}
\label{ChPT-together}
{\mathcal H}_{\rm ChPT}
= \frac{3}{4}f^2_{\eta^{\prime}}(\del_r\phi)^2 
+ \frac{3}{4r^2}f^2_{\eta^{\prime}} (\del_{\theta}\phi)^2 
+ \frac{3}{4}f^2_{\eta^{\prime}}(\del_z\phi)^2
+ a(1-{\rm cos} 3\phi)
- \frac{\mu^2_{\rm B}}{4\pi^2} \Omega \del_z\phi \,, 
\end{align}
where we ignored the terms of order ${\mathcal O}(\Omega^2)$ and we set ${\mathcal H}_{\rm ChPT} = 0$ in the QCD vacuum ($\phi=0$). To minimize the Hamiltonian (\ref{ChPT-together}), we need to set $\del_{\theta}\phi = 0$ and $\del_r\phi = 0$, and so the ground-state configuration is independent of $r$ and $\theta$. On the other hand, the effective Hamiltonian in the $z$ direction is 
\begin{gather}
{\mathcal H}_{\rm ChPT}
= \frac{3}{4}f^2_{\eta^{\prime}}(\del_z\phi)^2 + a(1-{\rm cos} 3\phi)
- \frac{\mu^2_{\rm B}}{4\pi^2} \Omega \del_z\phi \,,
\label{hamiltonian_zdirection}
\end{gather}
whose ground-state configuration can have a nontrivial $z$ dependence. As we will look for a configuration where the first term is comparable to the third term in eq.~(\ref{hamiltonian_zdirection}), for which $p \sim \mu_{\rm B}^2 \Omega/(4\pi^2 f_{\eta'}^2)$, the effective theory is valid when $\mu_{\rm B}^2 \Omega/(16\pi^3 f_{\eta'}^3) \ll 1$.

We note that the Hamiltonian (\ref{hamiltonian_zdirection}), which corresponds to the forth line of table~\ref{tab:csl}, is mathematically equivalent to that in QCD at finite $\mu_{\rm B}$ and $\mu_{\rm I}$ under rotation in ref.~\cite{Huang:2017pqe} (see eq.~(3.7)), which corresponds to the third line of table~\ref{tab:csl}. More concretely, the former is obtained from the latter by the following replacement:
\begin{gather}
\label{replacement-hadron}
\phi \to 3\phi \,, \quad f^2_{\pi} \to \frac{f^2_{\eta^{\prime}}}{6} \,, \quad m^2_{\pi} \to \frac{6a}{f^2_{\eta^{\prime}}} \,, \quad 
\mu_{\rm I} \Omega \to \frac{1}{6} \mu_{\rm B} \Omega \,,
\end{gather}
see also eq.~(2.2) in ref.~\cite{Brauner:2016pko}, which corresponds to the second line of table~\ref{tab:csl}.
Therefore, by making use of the results, e.g., in ref.~\cite{Huang:2017pqe}, one can obtain the ground state in the present case---the $\eta'$ CSL. It should be emphasized, however, that the physics is somewhat different: while the $\pi_0$ meson becomes massive by the quark mass in ref.~\cite{Huang:2017pqe}, the $\eta^{\prime}$ meson becomes massive by the ${\rm U}(1)_{\rm A}$ anomaly here.

To make this paper self-contained, we will briefly summarize the derivation of the ground state for the Hamiltonian (\ref{hamiltonian_zdirection}). 
The equation of motion is
\begin{gather}
\label{ChPT-eta-CSL}
\del^2_z(3\phi) = \frac{6a}{f^2_{\eta^{\prime}}} \, {\rm sin}(3\phi) \,,
\end{gather}
which can be analytically solved by using the Jacobi's elliptic function as
\begin{gather}
{\rm cos} \frac{3\phi({\bar z})}{2} = {\rm sn}({\bar z}, k) \,,
\end{gather}
where $\bar z \equiv z\sqrt{6a}/(f_{\eta^{\prime}}k)$ is a dimensionless coordinate and $k$ ($0 \le k \le 1$) is the elliptic modulus. This solution has a period 
\begin{gather}
\ell = \frac{2f_{\eta^{\prime}}kK(k)}{\sqrt{6a}}\,, 
\end{gather}
where $K(k)$ is the complete elliptic integral of the first kind.

The free parameter $k$ is determined by minimizing the total energy of the system at fixed volume $V$ with respect to $k$ as
\begin{gather}
\label{ChPT-eta-minicondition}
\frac{E(k)}{k}
= \frac{\mu^2_{\rm B}\Omega}{8\pi f_{\eta^{\prime}}\sqrt{6a}}\,,
\end{gather}
where $E(k)$ is the complete elliptic integral of the second kind.
The inequality $E(k)/k \geq 1$ leads to the critical angular velocity
\begin{gather}
\label{critical-massless}
\Omega_{\eta^{\prime}}^0 = \frac{8\pi f_{\eta^{\prime}}\sqrt{6a}}{\mu_{\rm B}^2}\,.
\end{gather}

We can also show that, when $\Omega > \Omega_{\eta^{\prime}}^0$, the $\eta'$ CSL is energetically more stable than the QCD vacuum for $\mu_{\rm B}<m_{\rm N}$ and nuclear matter for $\mu_{\rm B} \approx m_{\rm N}$ with $m_{\rm N}$ the nucleon mass. One can show the former statement by writing the energy of each lattice per unit area, satisfying the minimization condition (\ref{ChPT-eta-minicondition}), as
\begin{gather}
\label{E/S}
\frac{{\cal E}}{S}
= \int_0^{\ell}{\rm d}z {\cal H}_{\rm ChPT} = 
\frac{2 f_{\eta^{\prime}} \sqrt{6a}}{3} \left(k - \frac{1}{k} \right) K(k) < 0 \,.
\end{gather}
One can also show the latter by computing the baryon number of each lattice per unit area as
\begin{gather}
\label{N_B/S}
\frac{N_{\rm B}}{S}= -\int_0^{\ell}{\rm d}z \frac{\del {\cal H}_{\rm ChPT}}{\del \mu_{\rm B}} = \frac{\mu_{\rm B}\Omega}{3 \pi}\,,
\end{gather}
which, combined with eq.~(\ref{E/S}), indicates that the energy of the $\eta'$ CSL per unit baryon number is smaller than that of nuclear matter, $m_{\rm N} - \mu_{\rm B} \approx 0$.

\subsection{Ground state with finite quark masses}
\label{subsec:GS(finite quark masses)}
So far, we have ignored the explicit chiral symmetry breaking by the finite quark masses. We now discuss the quark mass effects on the $\eta'$ CSL. We assume that
\begin{gather}
\label{quark-mass@realistic}
m_{\rm u} = m_{\rm d} \equiv m_{\rm ud} \, \,, \quad m_{\rm ud} < m_{\rm s} \,,
\end{gather}
and that the quark mass term (\ref{mass-ChPT}) is sufficiently small compared with the ${\rm U}(1)_{\rm A}$ anomaly (\ref{ChPT-det-anomaly}) so that we can treat eq.~(\ref{mass-ChPT}) as a perturbation to eq.~(\ref{ChPT-det-anomaly}). Under this assumption, the off-diagonal parts of the mass matrix for $\eta$ and $\eta^{\prime}$ are negligible.
In this case, we can determine the ground state analytically even away from the chiral limit.

From the effective Hamiltonian
\begin{gather}
{\mathcal H}_{\rm ChPT}
=\frac{3}{4}f^2_{\eta^{\prime}}(\del_z\phi)^2 + a(1-{\rm cos} 3\phi) + 2B\tr M \, (1-{\rm cos}\phi)
-\frac{\mu^2_{\rm B}}{4\pi^2}\Omega \del_z\phi \,,
\end{gather}
we get the equation of motion
\begin{gather}
\del^2_z\phi
= A_1 {\rm sin}\phi + 3A_2 {\rm sin} 3\phi \label{EOM-mass} \,,
\end{gather}
where we set
\begin{gather}
A_1 \equiv \frac{4B\tr M}{3f^2_{\eta^{\prime}}}\,, \qquad A_2 \equiv \frac{2a}{3f^2_{\eta^{\prime}}}\,.
\end{gather}
One can solve this equation analytically in a similar way to solving the equation of motion for a single pendulum. 
This equation has a conserved quantity $C$,%
\footnote{There is an ambiguity on the choice of $C$: e.g., one could also choose $C' \equiv C - A_1$ as a conserved quantity.
The present choice will be convenient, as $C$ satisfies $C \geq A_1$ and so $\bar{k}$ defined below satisfies 
$\bar{k} \leq 1$ similarly to the elliptic modulus $k$ satisfying $k \leq 1$.}
\begin{gather}
\label{conserved energy}
C = \frac{1}{2}(\del_z\phi)^2 + A_1 {\rm cos}\phi + A_2 ({\rm cos}3\phi-1) \,.
\end{gather}
We then get
\begin{gather}
\frac{{\rm d}\phi}{{\rm d}z}
= \pm \frac{2\sqrt{A_1}}{\bar{k}} \left[
1 + \bar{k}^2\frac{A_2}{2A_1}(1-{\rm cos}3\phi) - \bar{k}^2{\rm cos}^2\frac{\phi}{2}
\right]^{\frac{1}{2}} \label{delphi-mass} \,,
\end{gather}
where $\bar{k} \equiv \sqrt{2A_1/(C+A_1)}$ is a counterpart of the elliptic modulus $k$ in eq.~(\ref{ChPT-eta-CSL}) and satisfies $0 < \bar{k} \le 1$. 
Without loss of generality, we can choose $\frac{{\rm d}\phi}{{\rm d}z} > 0$. Integrating eq.~(\ref{delphi-mass}) and taking $\phi(0)=0$, we have
\begin{gather}
\frac{2\sqrt{A_1}}{\bar{k}}z
= \int_{0}^{\phi}{\rm d}\theta \, \left[
1 + \bar{k}^2\frac{A_2}{2A_1}(1-{\rm cos}3\theta) - \bar{k}^2{\rm cos}^2\frac{\theta}{2}
\right]^{-\frac{1}{2}} \label{conf eta'-mass} \,.
\end{gather}
Unlike the conventional CSL state \cite{kishine2015theory}, this solution cannot be expressed by the Jacobi elliptic function.
Still, one can show that this solution has the following three properties. 
First, it is periodic in $\phi$ with the period
\begin{gather}
\bar{\ell} \equiv \frac{\bar{k} \bar{K}(\bar{k})}{2\sqrt{A_1}} \,, \qquad
\bar{K}(\bar{k}) \equiv \int_{0}^{2\pi}{\rm d}\theta \, \left[
1 + \bar{k}^2\frac{A_2}{2A_1}(1-{\rm cos}3\theta) - \bar{k}^2{\rm cos}^2\frac{\theta}{2}
\right]^{-\frac{1}{2}} \,.
\label{K_tilde}
\end{gather}
Second, it has a topological charge defined as 
\begin{gather}
\int_{0}^{\bar \ell}\frac{{\rm d}z}{2\pi} \, \del_z\phi(z) = \frac{1}{2\pi}[\phi(\bar{\ell}) - \phi(0)] = 1 \,,
\end{gather}
which depends only on the boundary values of each lattice. 
Third, it breaks parity symmetry because $\eta'$ is a pseudoscalar.

In summary, the ground state (\ref{conf eta'-mass}) is a periodic array of topological solitons which spontaneously breaks parity and continuous translational symmetries.
Since these properties are characteristic of the CSL, this ground state can be regarded as the CSL,  
although the ground-state configuration is mathematically different from the conventional CSL solution \cite{kishine2015theory}.
One can further show that this inhomogeneous state is energetically more favorable than the QCD vacuum ($\phi=0$) for $\Omega > \Omega_{\eta^{\prime}}$ (see appendix~\ref{app:field-conf}):
\begin{gather}
\Omega_{\eta^{\prime}} = \frac{6\pi f^2_{\eta^{\prime}}}{\mu^2_{\rm B}}
\int_{0}^{2\pi}{\rm d}\theta \, \left(
A_1 {\rm sin}^2\frac{\theta}{2} 
+ A_2 {\rm sin}^2\frac{3\theta}{2}
\right)^{\frac{1}{2}}
\label{critical ang(finite masses)} \,.
\end{gather}

Let us compare this critical angular velocity for the $\eta'$ CSL with that of the $\pi_0$ CSL at finite $\mu_{\rm B}$ and $\mu_{\rm I}$ previously derived in ref.~\cite{Huang:2017pqe}:
\begin{gather}
\Omega_{\pi_0}
= \frac{8\pi m_{\pi}f^2_{\pi}}{\mu_{\rm B}|\mu_{\rm I}|} \,. \label{critical ang_pion}
\end{gather}
One should notice the difference of the chemical potential dependences between the two: $\Omega_{\pi_0} \propto 1/(\mu_{\rm B} |\mu_{\rm I}|)$ while $\Omega_{\eta^{\prime}} \propto 1/\mu^2_{\rm B}$. In particular, when $\mu_{\rm B} \gg |\mu_{\rm I}|$, we have $\Omega_{\pi_0} \gg \Omega_{\eta^{\prime}}$, showing that the $\eta'$ CSL is realized much earlier than the $\pi_0$ CSL as $\Omega$ is increased.

When $\Omega > \Omega_{\pi_0}$, one needs to compare the total energy of the $\eta^{\prime}$ CSL with that of the $\pi_0$ CSL to determine which state is realized as a ground state. Such a comparison is feasible when $\Omega \approx \Omega_{\pi_0}$: the total energy of the former is given by eq.~(\ref{energy of eta' (finite mass)}) and satisfies $\bar{\mathcal E}_{\rm tot} < 0$. On the other hand, the total energy of the latter is ${\mathcal E}_{\rm tot}^{\pi_0} \approx 0$ \cite{Huang:2017pqe}. Therefore, the ground state in this case is the $\eta'$ CSL.

\section{High-density color-flavor locked (CFL) phase under rotation}
\label{sec:csl}
In this section, we show that the $\eta^{\prime}$ CSL also appears in the high-density CFL phase under rotation. 
Below we will consider the regime $\mu_{\rm B} \gg \Omega$.

\subsection{Effective theory of the CFL}
\label{subsec:EFT-CFL}
We first construct the low-energy effective theory for the CFL phase under rotation. The CFL phase is characterized by the diquark condensates \cite{Alford:1998mk}, 
\begin{gather}
\label{3flavor-condensate}
\langle q^{j}_{{\rm L} \beta}C q^{k}_{{\rm L} \gamma} \rangle 
= \epsilon^{ijk}\epsilon_{\alpha \beta \gamma}[d^{\dag}_{\rm L}]_{\alpha i} \,, \qquad
\langle q^{j}_{{\rm R} \beta}C q^{k}_{{\rm R} \gamma} \rangle
= \epsilon^{ijk}\epsilon_{\alpha \beta \gamma}[d^{\dag}_{\rm R}]_{\alpha i} \,,
\end{gather}
where $(i,j,k)$ and $(\alpha, \beta, \gamma)$ indicate flavor and color indices, respectively, and $C$ is the charge conjugation operator. These condensates lead to 10 NG bosons: the octet of mesons ($\tilde \pi, \tilde K, \tilde \eta$) associated with the spontaneous chiral symmetry breaking ${\rm SU}(3)_{\rm c} \times {\rm SU}(3)_{\rm R} \times {\rm SU}(3)_{\rm L} \to {\rm SU}(3)_{\rm c + R + L}$, and the H boson and $\tilde \eta^{\prime}$ meson associated with the spontaneous breaking of the ${\rm U}(1)_{\rm B}$ and ${\rm U}(1)_{\rm A}$ symmetries, respectively. (Here and below, the NG bosons in the CFL phase are denoted by $\tilde \pi, \tilde K, \tilde \eta$, etc. to distinguish from the mesons $\pi, K, \eta$, etc. in the hadronic phase.) Note that, due to the suppression of the instanton effect at large $\mu_{\rm B}$, this system has the approximate ${\rm U}(1)_{\rm A}$ symmetry and we can also regard the $\tilde \eta^{\prime}$ meson as the NG boson \cite{Son:2000fh}. Since the dynamics of the H boson is decoupled from other NG modes that we are interested in, we will ignore the H boson in the following discussion.

The nonet of the mesons corresponds to phase fluctuations defined by
\begin{gather}
d_{\rm L}d^{\dag}_{\rm R} = |d_{\rm L}d^{\dag}_{\rm R}|\, \tilde{U} \,,
\end{gather}
where the $\U(3)$ matrix $\tilde{U}$ is the field of the nonet of the mesons. We split $\tilde{U}$ into the octet and singlet parts as
\begin{gather}
\tilde{U} = \tilde{\Sigma}\, \exp\left(\frac{\rmi \lambda_0\tilde{\eta}^{\prime}}{f_{\tilde{\eta}^{\prime}}}\right) \,, \qquad 
\tilde{\Sigma} = {\rm exp} \left(\frac{{\rm i} \lambda_A\tilde{\pi}_A}{f_{\tilde \pi}}\right) \,.
\end{gather}
The field $\tilde{\Sigma}$ transforms under $\SU(3)_{\rm R} \times \SU(3)_{\rm L} \times \SU(3)_{\rm c}$ as
\begin{gather}
\tilde{\Sigma} \to V_{\rm L} \tilde{\Sigma} V_{\rm R}^{\dag} \,,
\end{gather}
while the $\tilde{\eta}^{\prime}$ meson transforms under the ${\rm U}(1)_{\rm A}$ as
\begin{gather}
\tilde{\eta}^{\prime} \to \tilde{\eta}^{\prime} + 4f_{\tilde \eta^{\prime}}\theta_0 \label{transform-eta-CFL} \,,
\end{gather}
where the factor 4 originates from the fact that the $\tilde \eta'$ meson in the CFL phase is a ${\bar q}{\bar q}qq$ state rather than a ${\bar q}q$ state as in the hadronic phase.

Setting $\tilde{\phi} \equiv \sqrt{6}\tilde{\eta}^{\prime}/(3f_{\tilde{\eta}^{\prime}})$, the kinetic terms for these NG modes up to ${\mathcal O}(\del^2)$ are \cite{Son:1999cm}
\begin{gather}
\label{kin-CFL} 
{\mathcal L}_{\rm kin}
= \frac{1}{4} f^2_{\tilde{\pi}} v_{\tilde \pi}^2 g^{\mu \nu}_{\tilde \pi} {\rm tr} 
\left(\del_{\mu}^{\tilde \pi} \tilde{\Sigma} \del_{\nu}^{\tilde \pi} \tilde{\Sigma}^{\dag} \right)
+ \frac{3}{4}f^2_{\tilde{\eta}^{\prime}} v_{\tilde \eta'}^2 g^{\mu \nu}_{\tilde{\eta}^{\prime}} \del_{\mu}^{\tilde \eta'} \tilde{\phi} \del_{\nu}^{\tilde \eta'} \tilde{\phi}  \,,
\end{gather}
where $f_{\tilde{\pi}, \tilde{\eta}^{\prime}}$ and $v_{\tilde{\pi}, \tilde{\eta}^{\prime}}$ are the decay constants and velocities of the octet and singlet mesons, respectively, 
and we defined $\del_{\mu}^{\tilde \pi}$ and $\del_{\mu}^{\tilde \eta'}$, such that $\del_{0}^{\tilde \pi, \tilde \eta'} \equiv (1/v_{\tilde \pi, \tilde \eta'}) \del_0$ and $\del_{i}^{\tilde \pi, \tilde \eta'} \equiv \del_i$. We also defined $g^{\mu \nu}_{\tilde{\pi}, \tilde{\eta}^{\prime}}$ as the inverses of the ``effective metrics'' $g_{\mu \nu}^{\tilde{\pi}, \tilde{\eta}^{\prime}}$ given by eq.~(\ref{metric-rot}) with the replacement $\Omega \rightarrow \Omega/v_{\tilde \pi, \tilde \eta'}$.
At sufficiently high density, $f_{\tilde{\pi}, \tilde{\eta}^{\prime}}$ and $v_{\tilde{\pi}, \tilde{\eta}^{\prime}}$ can be computed by the weak-coupling analysis as \cite{Son:1999cm}
\begin{gather}
f^2_{\tilde{\pi}} = \frac{21-8{\rm ln}2}{18}\frac{\mu^2_{\rm B}}{18 \pi^2} \,, \qquad
f^2_{\tilde{\eta}^{\prime}} = \frac{\mu^2_{\rm B}}{24 \pi^2} \label{decay const_high dense} \,, \\
v^2_{\tilde{\pi}} = v^2_{\tilde{\eta}^{\prime}} = \frac{1}{3} \,.
\end{gather}
As the energy scale of this effective theory must be much smaller than the mass gap of quarks (i.e., the superconducting gap $\Delta$), the expansion parameter of the effective theory is $p/\Delta \ll 1$ with $p$ being the momentum. We note that, due to the relation $\Delta \ll \mu_{\rm B}$ at sufficiently high density, the condition $p/(4 \pi f_{\tilde \pi, \tilde \eta'}) \sim p/\mu_{\rm B} \ll 1$ is then automatically satisfied.

Let us then turn to the potential terms for the NG modes which generally consist of three parts: the quark mass term, the instanton-induced term related to the $\U(1)_{\rm A}$ anomaly, and the topological term induced by rotation. 

First, we consider the quark mass term. Similarly to the derivation of eq.~(\ref{mass-ChPT}), this term can be written down based on the symmetries up to ${\mathcal O}(M^2)$ as \cite{Son:1999cm,Beane:2000ms}
\begin{gather}
\label{mass-CFL}
{\mathcal L}_{\rm mass}^{\rm CFL}
= c\left[{\rm det}M \, {\rm tr}(M^{-1} \tilde{U}) + {\rm h.c.}\right] \,.
\end{gather}
The form of this term is different from eq.~(\ref{mass-ChPT}) because the transformation law (\ref{transform-eta-CFL}) is different from eq.~(\ref{chiral-rotation-NG}). The parameter $c$ can be determined at sufficiently high density by the weak-coupling calculation as \cite{Son:1999cm,Son:2000tu}
\begin{gather}
c = \frac{3\Delta^2}{2\pi^2} \,.
\end{gather}

Next, we construct the potential term due to the $\U(1)_{\rm A}$ anomaly or the instanton effect. For this purpose, it is important to recall that the $\U(1)_{\rm A}$ anomaly induces a nonvanishing chiral condensate in the CFL phase even in the chiral limit. We define the chiral condensate as $\Phi_{ij} = \braket{{\bar q}_{\rm R}^j q_{\rm L}^i}$. Considering the transformation of $\Phi$ under $\SU(3)_{\rm R} \times {\rm SU}(3)_{\rm L} \times {\rm SU}(3)_{\rm c} \times \U(1)_{\rm A}$, 
\begin{gather}
\Phi \to {\rm e}^{-2{\rm i}\theta_0\lambda_0} V_{\rm L} \Phi V^{\+}_{\rm R},
\end{gather}
the effective potential at lowest order in $\Phi$ is given by \cite{Hatsuda:2006ps,Yamamoto:2007ah}
\begin{gather}
\label{1-inst chiral limit}
{\mathcal L}_{1\text{-inst}} = -\gamma \tr \left(\Phi^{\+} \tilde{U} + {\rm h.c.} \right) \,. 
\end{gather}
This term explicitly breaks ${\rm U}(1)_{\rm A}$ down to ${\mathbb Z}_6$ and it stems from the $\U(1)_{\rm A}$ anomaly. In fact, the parameter $\gamma$ can be explicitly computed from the instanton-induced interaction as (see appendix \ref{app:instanton-induced potential} for the detail)
\begin{gather}
\label{inst-gamma}
\gamma = \frac{18\pi^2 C_{N_{\rm c}, N_{\rm f}}}{(N_{\rm c}+1)^2} N_{\rm c}^{b+1} N_{\rm f}^{-\frac{b+5}{2}} \Gamma\left(\frac{b+5}{2}\right) 
\left(\frac{8\pi^2}{g^2} \right)^{\! 2N_{\rm c}+1} \left(\frac{\Lambda_{\rm QCD}}{\mu_{\rm B}}\right)^{\! b}\frac{\Delta^2}{\mu_{\rm B}} \,,
\end{gather}
where $N_{\rm f}=3$ is the number of flavors, $\Lambda_{\rm QCD}$ is the QCD scale, and
\begin{gather}
C_{N_{\rm c}, N_{\rm f}} = \frac{0.466 \exp(-1.679N_{\rm c})1.34^{N_{\rm f}}}{(N_{\rm c}-1)!(N_{\rm c}-2)!} \,, \\
b = \frac{11}{3}N_{\rm c} - \frac{2}{3}N_{\rm f} \,.
\label{C_Nc}
\end{gather}
Here we ignored the effects of rotation on the instantons, which is a reasonable approximation in the regime $\mu_{\rm B} \gg \Omega$.
We also ignored the multi-instanton contributions because instantons are dilute at sufficiently large $\mu_{\rm B}$. 

To calculate the chiral condensate $\Phi_{ij}$ in the chiral limit, we turn on a small quark mass for a moment, 
which will be turned off at the end of the computation.
The effective Lagrangian induced by the single instanton is \cite{Son:2000fh,Schafer:2002ty,Yamamoto:2008zw}
\begin{gather}
\label{1inst-mass}
{\cal L}_{\rm 1\mathchar`-inst}^{\rm CFL}
= \frac{\tilde{a}}{2}\left[{\rm tr}(M^{\dag}\tilde{U})+ {\rm h.c.} \right] + {\mathcal O}(M^2) \,,
\end{gather}
where $\tilde{a}$ denotes the strength of the ${\rm U}(1)_{\rm A}$ anomaly in the CFL phase, which can be determined 
through the instanton-induced six-fermion interaction as \cite{Schafer:2002ty,Yamamoto:2008zw}
\begin{gather}
\label{parameter-a}
\tilde{a} = \frac{24 C_{N_{\rm c}, N_{\rm f}}}{N^2_{\rm c}-1} N_{\rm c}^{b-1}
N_{\rm f}^{-\frac{b+3}{2}} \Gamma \left(\frac{b+3}{2}\right)
\left(\frac{8\pi^2}{g^2}\right)^{\! 2N_{\rm c}+1} \left(\frac{\Lambda_{\rm QCD}}{\mu_{\rm B}}\right)^{\! b} \mu_{\rm B} \Delta^2 \,.
\end{gather}
From the effective potential for the Lagrangian (\ref{1inst-mass}), 
$V_{\rm 1\mathchar`-inst}^{\rm CFL} = \left. - {\cal L}_{\rm 1\mathchar`-inst}^{\rm CFL} \right|_{\Sigma \rightarrow 1}$, 
the chiral condensate reads \cite{Yamamoto:2008zw}
\begin{gather}
\label{chiral cond}
\left.
\Phi_{ij} = \frac{\del V_{\rm 1\mathchar`-inst}^{\rm CFL}}{\del (M^{\+})_{ji}} \right|_{m \rightarrow 0}
= -\frac{\tilde{a}}{2}\delta_{ij} \,,
\end{gather}
which is nonvanishing even in the chiral limit. Then, eq.~(\ref{1-inst chiral limit}) becomes
\begin{gather}
\label{inst-CFL-potential chiral limit}
{\mathcal L}_{1\text{-inst}}
= \frac{1}{2}\gamma \tilde{a} \tr(\tilde{U}+\tilde{U}^{\+}) \,.
\end{gather}

Finally, let us write down the topological term in the CFL phase under rotation. Since the transport coefficient of the CVE is again topologically protected independently of symmetry breaking patterns, we can also use the formula (\ref{anom_cve}) in this case (but with a minor modification as we will immediately explain below), which yields
\begin{gather}
\label{cve-topo-CFL}
{\mathcal L}_{\rm CVE}
= \frac{\mu_{\rm B}^2}{8\pi^2f_{\tilde{\eta}^{\prime}}} \sqrt{\frac{2}{3}} {\bm \nabla} \tilde \eta^{\prime} \cdot {\bm \Omega} \,.
\end{gather}
Note that the coefficient here is one half of that of eq.~(\ref{eta-cve-topological}) because of the difference of the transformation laws (\ref{chiral-rotation-NG}) and (\ref{transform-eta-CFL}) under the ${\rm U}(1)_{\rm A}$ rotation.

In summary, the effective theory of the CFL phase under rotation is given by eqs.~(\ref{kin-CFL}), (\ref{mass-CFL}), (\ref{1-inst chiral limit}), and (\ref{cve-topo-CFL}). 
We are now ready to consider the ground state of this system.

\subsection{Ground state in the chiral limit}
We first consider the case in the chiral limit. Similarly to the previous discussion on the low-density hadronic matter, we will focus on $\tilde \eta'$. Adding the eqs.~(\ref{kin-CFL}), (\ref{inst-CFL-potential chiral limit}) and (\ref{cve-topo-CFL}) together, we obtain the effective Hamiltonian for $\tilde \eta'$,
\begin{gather}
{\mathcal H}_{\rm CFL}
= \frac{\mu^2_{\rm B}}{96\pi^2}(\del_z\tilde{\phi})^2
+ 3\gamma \tilde{a} (1-{\rm cos}\tilde{\phi})
-\frac{\mu^2_{\rm B}}{8\pi^2}\Omega \del_z\tilde{\phi} \,.
\end{gather}
As in the case of low-density hadronic matter in section \ref{subsec:GSofChPT}, we will look for a ground-state solution where the first term is comparable to the third term, which requires that $\Omega/\Delta \ll 1$ for the effective theory to be valid.
This Hamiltonian is again mathematically equivalent to that of QCD at finite $\mu_{\rm B}$ and $\mu_{\rm I}$ under rotation in ref.~\cite{Huang:2017pqe}:
the latter is mapped to the former by the replacement
\begin{gather}
\label{eft-CFL}
f^2_{\pi} \to \frac{\mu_{\rm B}^2}{48\pi^2} \, \,, \quad 
m^2_{\pi} \to \frac{144\pi^2\tilde{a}\gamma}{\mu_{\rm B}^2} \, \,, \quad 
\mu_{\rm I} \Omega \to \frac{1}{4}\mu_{\rm B}\Omega \,.
\end{gather}

By repeating the discussion in section \ref{subsec:GSofChPT}, we find that the ground state is the CSL of the $\tilde \eta'$ meson, whose configuration is given by
\begin{gather}
\label{CSL@CFL}
{\rm cos} \frac{\tilde{\phi}({\bar z})}{2} = {\rm sn}({\bar z}, k) \,, \qquad {\bar z} \equiv \frac{12 \pi \sqrt{\gamma \tilde a}}{\mu_{\rm B}k}z \,.
\end{gather}
This solution has a period 
\begin{gather}
\tilde{\ell} = \frac{\mu_{\rm B} k K(k)}{6\pi \sqrt{\gamma \tilde a}}\,,
\end{gather}
and the critical angular velocity is
\begin{gather}
\label{mini-condition_CFL}
\Omega_{\tilde{\eta}^{\prime}}^0 = \frac{8\sqrt{\gamma \tilde{a}}}{\mu_{\rm B}} \,,
\end{gather}
above which the CSL state is energetically favorable than the CFL state ($\tilde{\phi}=0$). The $\mu_{\rm B}$ dependence of $\Omega_{\tilde{\eta}^{\prime}}^0$ is determined by substituting eqs.~(\ref{inst-gamma}) and (\ref{parameter-a}) into eq.~(\ref{mini-condition_CFL}) as
\begin{gather}
\label{critical ang CFL}
\Omega_{\tilde \eta^{\prime}}^0
=2^53^{\frac{11}{2}}5 \pi C_{N_{\rm c}, N_{\rm f}} \left(\frac{8\pi^2}{g^2}\right)^{\! 7} 
\left(\frac{\Lambda_{\rm QCD}}{\mu_{\rm B}}\right)^{\! 10}\frac{\Delta^2}{\Lambda_{\rm QCD}} \,.
\end{gather}
In particular, $\Omega_{\tilde \eta^{\prime}}^0 \rightarrow 0$ at asymptotically large $\mu_{\rm B}$.

\subsection{Ground state with finite quark masses}
Let us now turn on finite quark masses in flavor-symmetric QCD. 
In the case with flavor asymmetry, the equation of motion of the system is not analytically solvable even without 
the instanton-induced term and the CSL-type solution does not exist (see appendix~\ref{app:CFL-asym}).
Below we consider sufficiently large $\mu_{\rm B}$ for simplicity, where the instanton-induced term (\ref{1inst-mass}) is 
suppressed compared with the quark mass term in eq.~(\ref{mass-CFL}).

Let us consider eq.~(\ref{mass-CFL}) with the flavor symmetric masses:
\begin{gather}
M = {\rm diag}(m, m, m) \,.
\end{gather}
In this case, there is no mixing between $\tilde{\eta}^{\prime}$ meson and the other mesons in the mass matrix. 
Then, we obtain the effective Hamiltonian
\begin{gather}
{\mathcal H}_{\rm CFL}
= \frac{\mu_{\rm B}^2}{96\pi^2} (\del_z\tilde{\phi})^2 + \frac{9m^2 \Delta^2}{\pi^2}(1 - {\rm cos}\tilde{\phi}) 
- \frac{\mu_{\rm B}^2}{8\pi^2}\Omega \del_z\tilde{\phi} \,,
\end{gather}
Similarly to the discussions in previous sections, the ground state of this Hamiltonian is obtained from eq.~(3.9) in ref.~\cite{Huang:2017pqe} by the following replacement:
\begin{gather}
\label{eft-CFL}
f^2_{\pi} \to \frac{\mu_{\rm B}^2}{48\pi^2} \, \,, \quad 
m^2_{\pi} \to \frac{432 m^2 \Delta^2}{\mu_{\rm B}^2} \, \,, \quad 
\mu_{\rm I} \Omega \to \frac{1}{4}\mu_{\rm B}\Omega \,.
\end{gather}
In particular, the critical angular velocity for the CSL state is found as
\begin{gather}
\label{critical ang(finite masses) CFL}
\Omega_{\tilde{\eta}^{\prime}}
= \frac{8\sqrt{3}m\Delta}{\pi\mu_{\rm B}} \,.
\end{gather}

\section{Discussion and conclusion} 
\label{sec:discussion}
We have shown that the ground state of (nearly) three-flavor symmetric QCD under sufficiently fast rotation is the $\eta^{\prime}$ CSL both in low-density hadronic matter and high-density CFL phase. The critical angular velocity for the $\eta^{\prime}$ CSL is given in eq.~(\ref{critical ang(finite masses)}) at low density and eq.~(\ref{critical ang(finite masses) CFL}) at high density. In both regions, the parity symmetry and the continuous translational symmetry in the direction of the angular velocity are spontaneously broken, leading to a phonon as the additional low-energy excitation. Since the symmetry breaking patterns of the ground states and excitations around them are the same between the two regions, it is plausible that the $\eta'$ CSLs on both sides are continuously connected. This can be regarded as an extension of the quark-hadron continuity conjecture \cite{Schafer:1998ef, Hatsuda:2006ps,Yamamoto:2007ah} in the presence of rotation. On the other hand, in the case away from the flavor symmetry where the quark mass difference can no longer be treated as a perturbation, we cannot analytically solve the equation of motion for $\eta'$, and as a result, the ground state can be different from the $\eta'$ CSL.

In our analysis, we have studied the leading-order low-energy effective theory, where the interactions between $\eta'$ and other mesons (including the superfluid phonon associated with the spontaneous $\U(1)_{\rm B}$ symmetry breaking) are higher order in derivatives and are negligible. In this context, it should be remarked that, in the case of the $\pi_0$ CSL in baryonic matter with an external magnetic field ${\bm B}$, the term $\sim \del^{\mu} \pi_0 (\pi_1 \del_{\mu} \pi_2 - \pi_2 \del_{\mu} \pi_1)$ in the presence of the $\pi_0$ CSL background $\langle {\bm \nabla}\pi_0 \rangle \sim \mu_{\rm B} {\bm B}$ can lead to Bose-Einstein condensation of charged pions when the magnitude of the magnetic field is increased  \cite{Brauner:2016pko}. However, such a phenomenon will not occur for the $\eta'$ CSL in baryonic matter when $\Omega$ is increased further, since the kinetic term for $\eta'$ is decoupled from those of other mesons at the leading order. We hence expect that the $\eta'$ CSL is always realized above the critical angular velocity within the applicability of our effective theory at leading order. It would be interesting to study the possible modifications of our results due to the next-to-leading order corrections and the flavor asymmetry. 

Finally, let us discuss a possible realization of the rotation-induced $\eta^{\prime}$ CSL in physical systems. One possible candidate is noncentral heavy ion collision experiments. There, we may roughly estimate $\Omega_{\eta^{\prime}}$ and $\Omega_{\pi_0}$ for a rotating nuclear matter made up from $^{197}_{\ 79}$Au with saturation density $n \approx 0.16/{\rm fm}^3$ (corresponding to $\mu_{\rm B} \approx 1\, {\rm GeV}$ and $\mu_{\rm I} \approx -10\, {\rm MeV}$) as%
\footnote{Although the validity of the low-energy effective field theory for such large $\mu_{\rm B}$ is subtle, we here simply extrapolate the parametric dependence of $\Omega_{\eta^{\prime}}$ and $\Omega_{\pi_0}$ in eqs.~(\ref{critical ang(finite masses)}) and (\ref{critical ang_pion}) into such a regime to provide semi-quantitative estimates.}
\begin{gather}
\Omega_{\eta^{\prime}} \approx 4\times 10^2\, {\rm MeV}, \qquad
\Omega_{\pi_0} \approx 6\times 10^3\, {\rm MeV} \,,
\end{gather}
respectively, where we used the vacuum values of the quantities $f_{\pi} \approx 93\, {\rm MeV}$, $f_{\eta'}/f_{\pi} \approx 1.1$,  $m_{\pi} \approx 140$ MeV, $m_{\rm K} \approx 500$ MeV and $m_{\eta'} \approx 960$ MeV \cite{Donoghue:1992dd}, together with the Gell-Mann-Oakes-Renner relation
\begin{gather}
f_{\pi}^2 m_{\pi}^2 = 4B m _{\rm ud}, \qquad f_{\pi}^2 m_{\rm K}^2 = 2B (m _{\rm ud} + m_{\rm s}),
\end{gather}
and 
\begin{gather}
f_{\eta'}^2 m_{\eta'}^2 = \frac{4 B}{3} (2m _{\rm ud} + m_{\rm s}) + 6a\,.
\end{gather}
This result shows that $\Omega_{\eta^{\prime}}$ is an order of magnitude smaller than $\Omega_{\pi_0}$ previously found in ref.~\cite{Huang:2017pqe}, although it does not still reach the experimentally measured angular velocity in noncentral heavy ion collisions, $\Omega_{\rm exp} \sim 10\, {\rm MeV}$ \cite{STAR:2017ckg}. Nonetheless, because $\Omega_{\eta^{\prime}}$ decreases with increasing $\mu_{\rm B}$ as eqs.~(\ref{critical ang(finite masses)}) and (\ref{critical ang(finite masses) CFL}), the ${\eta}^{\prime}$ CSL could be potentially realized in high-density matter to be produced in low-energy heavy ion collision experiments in the future. (Even in such a case, it might be difficult to realize the $\pi_0$ CSL since $\Omega_{\pi_0} \gg \Lambda_{\rm QCD}$.)
To understand its possible realization in heavy ion collisions more realistically, it is necessary to take into account the finite-temperature effects on the $\eta'$ CSL; see also ref.~\cite{Brauner:2017mui} for the study on the low-temperature effects on the $\pi_0$ CSL in a magnetic field. Further studies in this direction will be reported elsewhere.

\acknowledgments
We are indebted to Xu-Guang Huang for collaboration preceding the present work. One of the authors (K.~N.) thanks N.~Ikoma and K.~Soga for discussions on the differential equation.
This work is supported by Keio Institute of Pure and Applied Sciences (KiPAS) project in Keio University and MEXT-Supported Program for the Strategic Research Foundation at Private Universities, ``Topological Science'' (Grant No.~S1511006). K.~N. is supported by JSPS KAKENHI Grant No.~19J21593. N.~Y. is supported by JSPS KAKENHI Grant No.~19K03852.

\appendix

\section{The $\eta^{\prime}$ potential in large-$N_{\rm c}$ QCD}
\label{app:Eta-largeN}
In this appendix, we show that the Lagrangian~(\ref{axial_anomaly}) explicitly breaks the continuous translational symmetry in space.

\begin{figure}[t]
\centering 
\includegraphics[width=10.0cm]{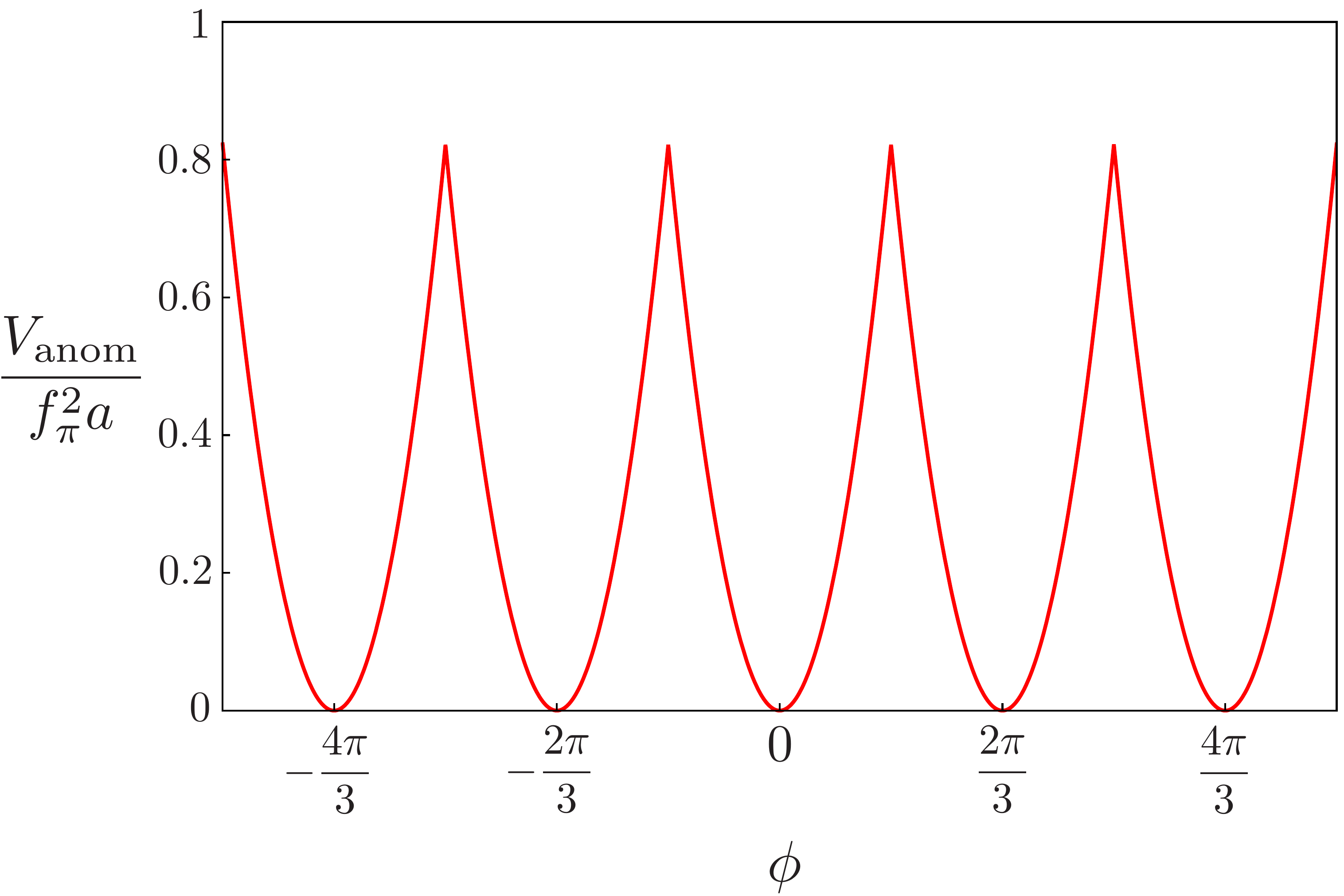}
\caption{\label{fig:ax} $V_{\rm anom}$ as a function of $\phi$.}
\end{figure}

Using the equality, 
${\rm det}\, {\rm exp}({\rm i}\lambda_a\pi_a/f_{\pi}) = {\rm exp} \, {\rm tr} ({\rm i}\lambda_a\pi_a/f_{\pi})$, eq.~(\ref{axial_anomaly}) becomes
\begin{align}
{\mathcal L}_{\rm anom}'
= -\frac{f^2_{\pi}a}{4N_{\rm c}} \left(3 \phi - 2\pi n \right)^{\! 2}\,,
\label{axial_anomaly2} 
\end{align}
where $\phi \equiv \sqrt{6}\eta^{\prime}/(3f_{\pi})$ as defined in the main text and $n$ is integer. The corresponding potential energy is
\begin{gather}
\label{axial_minimum}
V_{\rm anom}
= \frac{f_{\pi}^2 a}{4N_{\rm c}}\left(3\phi - 2\pi n \right)^2 \, \,, \quad 
(2n-1)\pi \le 3\phi < (2n+1)\pi \,,
\end{gather}
as sketched in figure~\ref{fig:ax}. Note that this potential has cusp singularities at $3 \phi = (2n-1)\pi$ with integer $n$.

\begin{figure}[t]
\centering
\includegraphics[width=17.0cm]{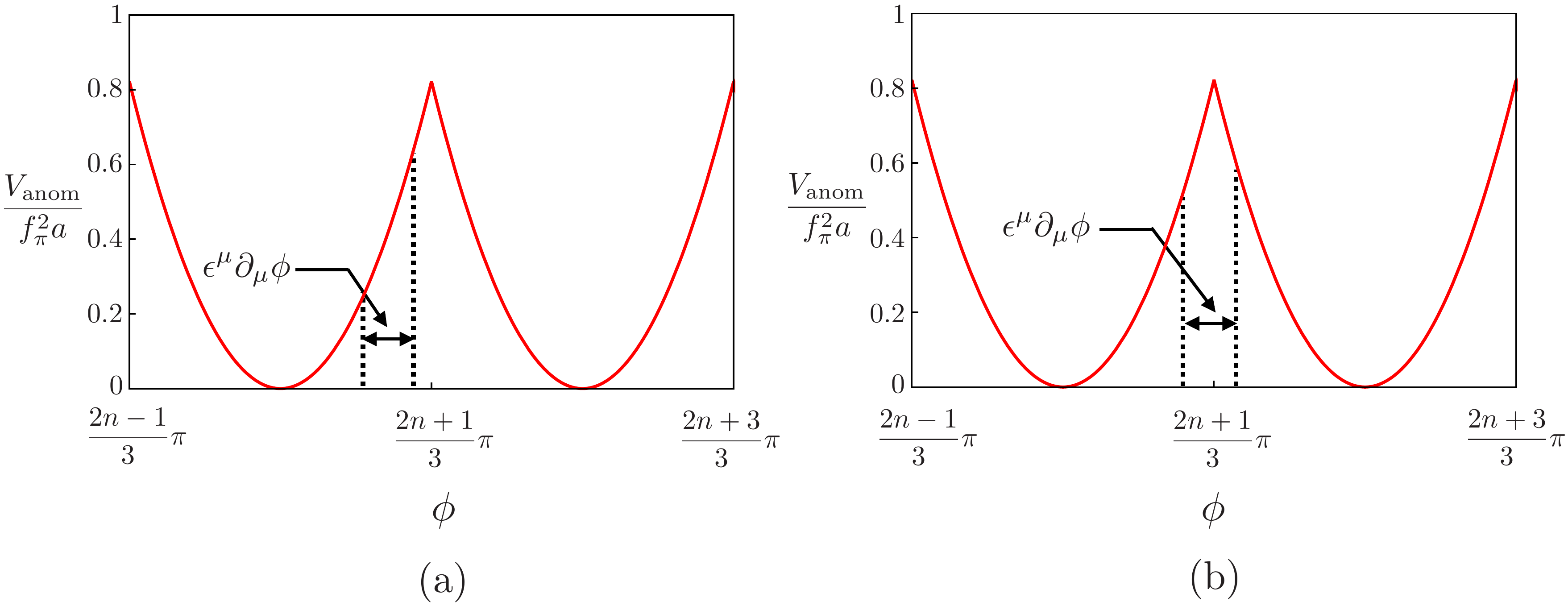}
\caption{\label{fig:translation-combination} (a) Both $\phi$ and $\phi+\delta \phi$ are between $(2n-1)\pi/3$ and $(2n+1)\pi/3$. (b) $\phi$ is between $(2n-1)\pi/3$ and $(2n+1)\pi/3$, while $\phi + \delta \phi$ is between $(2n+1)\pi/3$ and $(2n+3)\pi/3$.}
\end{figure}

To demonstrate the explicit breaking of spacial translational symmetry, consider an infinitesimal spacetime translation $x^{\mu} \to x^{\mu} - \epsilon^{\mu}$, under which the variation of $\phi$ is $\delta \phi(x) = \epsilon^{\mu}\del_{\mu}\phi(x) + {\mathcal O}(\epsilon^2)$. When $\delta \phi$ does not cross a cusp singularity, i.e., when $(2n-1)\pi/3 \le \phi \le (2n+1)\pi/3$ and $(2n-1)\pi/3 \le \phi+\delta \phi \le (2n+1)\pi/3$ (see figure \ref{fig:translation-combination}), the variation of eq.~(\ref{axial_anomaly}) is written as a total derivative:
\begin{align}
\delta_{\rm L}{\mathcal L}_{\rm anom}' = \epsilon^{\nu}\del_{\nu}{\mathcal L}_{\rm anom}' \,,
\end{align}
and thus, it is invariant under the spacetime translation. However, when $\delta \phi$ crosses a cusp singularity, i.e., when $(2n-1)\pi/3 \le \phi \le (2n+1)\pi/3$ and $(2n+1)\pi/3 \le \phi + \delta \phi \le (2n+3)\pi/3$ (see figure \ref{fig:translation-combination}), the variation of eq.~(\ref{axial_anomaly}) is 
\begin{align}
\delta_{\rm L}{\mathcal L}_{\rm anom}' = \epsilon^{\nu}\del_{\nu}{\mathcal L}_{\rm anom}'
+ \frac{3\pi f^2_{\pi}a}{N_{\rm c}}\left(\phi - \frac{2n + 1}{3}\pi
\right) \,,
\end{align}
where the second term is not a total derivative. Therefore, the Lagrangian (\ref{axial_anomaly}) is not invariant under the spatial translational symmetry.

\section{Derivation of the critical angular velocity $\Omega_{\eta^{\prime}}$}
\label{app:field-conf}
In this appendix, we give the derivation of $\Omega_{\eta^{\prime}}$ in eq.~(\ref{critical ang(finite masses)}).
The energy of each lattice per unit area is
\begin{align}
\frac{\bar{\cal E}}{S} = \frac{3}{2}f^2_{\eta^{\prime}}\sqrt{A_1}\left[
\frac{2}{\bar{k}}\bar{E}(\bar{k}) + \left(\bar{k}-\frac{1}{\bar{k}} \right)\bar{K}(\bar{k})
\right] - \frac{\mu^2_{\rm B}\Omega}{2\pi} \,,
\end{align}
where $\bar{K}(\bar{k})$ is defined in eq.~(\ref{K_tilde}) and $\bar{E}(\bar{k})$ is defined as
\begin{gather}
\bar{E}(\bar{k}) \equiv \int_{0}^{2\pi}{\rm d}\theta \, \left[
1 + \bar{k}^2\frac{A_2}{2A_1}(1-{\rm cos}3\theta) - \bar{k}^2{\rm cos}^2\frac{\theta}{2}
\right]^{\frac{1}{2}} \,, 
\end{gather}
both of which should not be confused with the complete elliptic integrals of the first and second kinds, $K(k)$ and $E(k)$.
The total energy of the system of length $L$ is given by
\begin{gather}
\bar{\cal E}_{\rm tot} \equiv \frac{L}{\bar{\ell}}\bar{\cal E} 
= 3V f^2_{\eta^{\prime}}A_1 \left[\frac{2 \bar E(\bar k)}{\bar{k}^2 \bar{K}(\bar{k})} + 1 - \frac{1}{\bar{k}^2} - \frac{\mu^2_{\rm B}\Omega}{3 \pi f^2_{\eta^{\prime}}\sqrt{A_1}} \frac{1}{\bar k \bar K (\bar k)} \right],
\end{gather}
where $\bar{\ell}$ is defined in eq.~(\ref{K_tilde}) and $V \equiv LS$.

To minimize $\bar{\cal E}_{\rm tot}$ with respect to $\bar{k}$, we calculate its derivative as
\begin{gather}
\frac{{\rm d}\bar{\cal E}_{\rm tot}}{{\rm d}\bar{k}}
= 3V f^2_{\eta^{\prime}}A_1\frac{\bar{H}(\bar{k})}{\bar{k}^2 \bar{K}(\bar{k})^2} \left[
\frac{\mu^2_{\rm B}\Omega}{3 \pi f^2_{\eta^{\prime}}\sqrt{A_1}} - \frac{2\bar{E}(\bar{k})}{\bar{k}}
\right] \label{derivative-energy} \,,
\end{gather}
where we used the relations
\begin{gather}
\frac{{\rm d} \bar{E}(\bar{k})}{{\rm d}\bar{k}} = \frac{\bar{E}(\bar{k})- \bar{K}(\bar{k})}{\bar{k}} \,, \\
\frac{{\rm d} \bar{K}(\bar{k})}{{\rm d}\bar{k}} = \frac{\bar{H}(\bar{k})- \bar{K}(\bar{k})}{\bar{k}} \,,
\end{gather}
and we defined
\begin{gather}
\bar{H}(\bar{k}) \equiv \int_{0}^{2\pi}{\rm d}\theta \, \left[1 + \bar{k}^2\frac{A_2}{2A_1}(1-{\rm cos}3\theta) - \bar{k}^2{\rm cos}^2\frac{\theta}{2} \right]^{-\frac{3}{2}} \,.
\end{gather}
Because the factor in front of the bracket in eq.~(\ref{derivative-energy}) is positive, we can focus on the factor inside the bracket, 
\begin{gather}
g(\bar{k}) \equiv \frac{\mu^2_{\rm B}\Omega}{3 \pi f^2_{\eta^{\prime}}\sqrt{A_1}} - \frac{2 \bar{E}(\bar{k})}{\bar{k}} \,.
\end{gather}
Since $\bar{E}(\bar{k})/\bar{k}$ monotonically decreases as a function of $\bar{k}$ and it is bounded from below as $\bar{E}(\bar{k})/\bar{k} \ge \bar{E}(1)$ for $0\le \bar{k} \le 1$, 
the CSL solution exists if and only if $\Omega \geq \Omega_{\eta^{\prime}}$, where $\Omega_{\eta^{\prime}}$ is given by eq.~(\ref{critical ang(finite masses)}).
We can also calculate $\bar{\cal E}_{\rm tot}$ satisfying the minimization condition ${\rm d} \bar{\cal E}_{\rm tot} /{\rm d} \bar{k}=0$ as
\begin{gather}
\label{energy of eta' (finite mass)}
\bar{\cal E}_{\rm tot} = 3V f^2_{\eta^{\prime}}A_1 \left(1-\frac{1}{\bar{k}^2}\right) < 0 \,.
\end{gather}
Therefore, the CSL solution is energetically more favorable than the QCD vacuum ($\phi=0$).

\section{Calculation of the instanton-induced potential}
\label{app:instanton-induced potential}
We here provide the derivation of eqs.~(\ref{1-inst chiral limit}) and (\ref{inst-gamma}).
We start from the six-fermion instanton-induced vertex \cite{tHooft:1976snw,Shifman:1979uw,Schafer:1996wv}
\begin{gather}
{\mathcal L}_{\rm inst}
= \int_{0}^{\infty}d\rho \, n(\rho) \frac{(2\pi \rho)^6\rho^3}{6N_{\rm c}(N^2_{\rm c}-1)} 
\epsilon_{i_1i_2i_3} \epsilon_{j_1j_2j_3} \Biggl[
\frac{2N_{\rm c}+1}{2N_{\rm c}+4} ({\bar q}_{\rm R}^{i_1}q_{\rm L}^{j_1}) ({\bar q}_{\rm R}^{i_2}q_{\rm L}^{j_2})
({\bar q}_{\rm R}^{i_3}q_{\rm L}^{j_3}) \notag \\
-\frac{3}{8(N_{\rm c}+2)} ({\bar q}_{\rm R}^{i_1}q_{\rm L}^{j_1})
({\bar q}_{\rm R}^{i_2}\sigma_{\mu \nu}q_{\rm L}^{j_2}) ({\bar q}_{\rm R}^{i_3}\sigma^{\mu \nu}q_{\rm L}^{j_3})
+ ({\rm R} \leftrightarrow {\rm L})
\Biggr] \label{inst-vertex} \,,
\end{gather}
where $\rho$ is the instanton size, $i_{1,2,3}$ and $j_{1,2,3}$ are flavor indices, and $\sigma_{\mu \nu} = \frac{\rm i}{2}[\gamma_{\mu}, \gamma_{\nu}]$. The instanton size distribution $n(\rho)$ is given by \cite{Schafer:1996wv,Shuryak:1982hk}
\begin{gather}
n(\rho) = C_{N_{\rm c}, N_{\rm f}} \left(\frac{8\pi^2}{g^2} \right)^{2N_{\rm c}} \rho^{-5} {\rm exp} \left(-\frac{8\pi^2}{g(\rho)^2} \right)
{\rm e}^{-N_{\rm f}\mu^2\rho^2} \,, \\
\frac{8\pi^2}{g(\rho)^2} = -b \ln(\rho \Lambda_{\rm QCD}) \,,
\end{gather}
where $C_{N_{\rm c}, N_{\rm f}}$ and $b$ are defined in eq.~(\ref{C_Nc}).

We evaluate the expectation value of eq.~(\ref{inst-vertex}) in the presence of the diquark and chiral condensate in the mean-field approximation. 
Replacing ${\bar q}_{\rm R}q_{\rm L}$ with the chiral condensate $\Phi$, we get
\begin{gather}
{\mathcal L}_{\rm inst}
\simeq \int_{0}^{\infty}d\rho \, n(\rho) \frac{(2\pi \rho)^6\rho^3}{6N_{\rm c}(N^2_{\rm c}-1)} 
\epsilon_{i_1i_2i_3} \epsilon_{j_1j_2j_3} \Biggl[
\frac{2N_{\rm c}+1}{2N_{\rm c}+4} 3({\bar q}_{\rm R}^{i_1}q_{\rm L}^{j_1}) ({\bar q}_{\rm R}^{i_2}q_{\rm L}^{j_2})
\Phi_{j_3i_3} \notag \\
-\frac{3}{8(N_{\rm c}+2)} \Phi_{j_1i_1}
({\bar q}_{\rm R}^{i_2}\sigma_{\mu \nu}q_{\rm L}^{j_2}) ({\bar q}_{\rm R}^{i_3}\sigma^{\mu \nu}q_{\rm L}^{j_3})
+ ({\rm R} \leftrightarrow {\rm L})
\Biggr] \,.
\end{gather}
To replace the remaining part with the diquark condensate, we use the Fierz transformations
\begin{gather}
(1-\gamma_5)_{\sigma \sigma'} (1-\gamma_5)_{\tau \tau'}
= -\frac{1}{2} \left[(1-\gamma_5)C \right]_{\sigma \tau} [C(1-\gamma_5)]_{\tau' \sigma'}
-\frac{1}{4} \left[(1-\gamma_5)\sigma_{\mu \nu}C \right]_{\sigma \tau}(C\sigma^{\mu \nu})_{\tau' \sigma'} \,, \\
[\sigma^{\mu \nu}(1-\gamma_5)]_{\sigma \sigma'}[\sigma_{\mu \nu}(1-\gamma_5)]_{\tau \tau'}
= 6[(1-\gamma_5)C]_{\sigma \tau} [C(1-\gamma_5)]_{\tau' \sigma'}
-[(1-\gamma_5)\sigma_{\mu \nu}C]_{\sigma \tau}(C\sigma^{\mu \nu})_{\tau' \sigma'} \,,
\end{gather}
and the diquark condensate (\ref{3flavor-condensate}), which is related to the superconducting gap $\Delta$ by \cite{Schafer:1999fe} 
\begin{gather}
|d_{\rm R}| = |d_{\rm L}| = \frac{3\sqrt{2}\mu_{\rm B}^2 \Delta}{2\pi gN^2_{\rm c}} \,.
\end{gather}
Then, we arrive at the instanton-induced potential in eqs.~(\ref{1-inst chiral limit}) and (\ref{inst-gamma}).

\section{CFL phase with flavor asymmetry}
\label{app:CFL-asym}
We here consider the CFL phase with flavor-asymmetric quark masses given in eq.~(\ref{quark-mass@realistic}). 
Expanding eq.~(\ref{mass-CFL}) to the second order in meson fields, we have a $9 \times 9$ real-symmetric mass matrix for them, which can be decomposed in to  $7 \times 7$ matrix for $\tilde \pi$ and $\tilde K$ and a nondiagonal $2 \times 2$ matrix for $\tilde{\eta}$ and $\tilde{\eta}^{\prime}$. As we are interested in the low-energy dynamics and the lightest meson in the CFL phase is a mixed state of $\tilde{\eta}$ and $\tilde{\eta}^{\prime}$ due to the inverse meson mass ordering \cite{Son:1999cm}, we will focus on the nondiagonal $2 \times 2$ matrix for $\tilde{\eta}$ and $\tilde{\eta}^{\prime}$ in the following.%
\footnote{See also ref.~\cite{Manuel:2000wm} for a related work on the meson mass spectra in the CFL phase. Note that we here set $m_{\rm u} = m_{\rm d}$ (see eq.~(\ref{quark-mass@realistic})), and in this case, $\tilde{\eta}$ and $\tilde{\eta}^{\prime}$ do not mix with $\tilde \pi_0$ in the mass matrix unlike refs.~\cite{Son:1999cm, Manuel:2000wm}.}

The elements of the mass matrix in the $\tilde{\eta}$-$\tilde{\eta}^{\prime}$ sector in eq.~(\ref{mass-CFL}) are
\begin{gather}
(M_{\tilde{\eta} \mathchar`- \tilde{\eta}^{\prime}})_{00}
= \frac{4cm_{\rm ud}(m_{\rm ud}+2m_{\rm s})}{3f^2_{\tilde{\eta}^{\prime}}} \,, \qquad
(M_{\tilde{\eta} \mathchar`- \tilde{\eta}^{\prime}})_{88}
= \frac{4cm_{\rm ud}(2m_{\rm ud}+m_{\rm s})}{3f^2_{\tilde{\pi}}} \,, \notag \\
(M_{\tilde{\eta} \mathchar`- \tilde{\eta}^{\prime}})_{08} = (M_{\tilde{\eta} \mathchar`- \tilde{\eta}^{\prime}})_{80}
= \frac{4\sqrt{2}cm_{\rm ud}(m_{\rm s}-m_{\rm ud})}{3f_{\tilde{\eta}^{\prime}}f_{\tilde{\pi}}} \,.
\label{mass-matrix}
\end{gather}
We can diagonalize it by the real-orthogonal matrix $T$ as
\begin{gather}
{}^tT M_{\tilde \eta \mathchar`- \tilde \eta^{\prime}} T
= \frac{1}{2} {\rm diag}\bigg((M_{\tilde{\eta} \mathchar`- \tilde{\eta}^{\prime}})_{00}+(M_{\tilde{\eta} \mathchar`- \tilde{\eta}^{\prime}})_{88}-\sqrt{\big[(M_{\tilde{\eta} \mathchar`- \tilde{\eta}^{\prime}})_{00}-(M_{\tilde{\eta} \mathchar`- \tilde{\eta}^{\prime}})_{88}\big]^2+4(M_{\tilde{\eta} \mathchar`- \tilde{\eta}^{\prime}})_{08}^2} \,, \notag \\
\qquad \qquad \qquad \qquad (M_{\tilde{\eta} \mathchar`- \tilde{\eta}^{\prime}})_{00}+(M_{\tilde{\eta} \mathchar`- \tilde{\eta}^{\prime}})_{88}+\sqrt{\big[(M_{\tilde{\eta} \mathchar`- \tilde{\eta}^{\prime}})_{00}-(M_{\tilde{\eta} \mathchar`- \tilde{\eta}^{\prime}})_{88}\big]^2+4(M_{\tilde{\eta} \mathchar`- \tilde{\eta}^{\prime}})_{08}^2} \bigg) \,,
\label{mass-eigenvalue}
\end{gather}
where the mixing angle $\theta$ satisfies
\begin{gather}
{\rm tan}\theta
= \frac{2(M_{\tilde \eta \mathchar`- \tilde \eta^{\prime}})_{08}}{-(M_{\tilde \eta \mathchar`- \tilde \eta^{\prime}})_{00}
+(M_{\tilde \eta \mathchar`- \tilde \eta^{\prime}})_{88}
+\sqrt{\big[(M_{\tilde \eta \mathchar`- \tilde \eta^{\prime}})_{00}-(M_{\tilde \eta \mathchar`- \tilde \eta^{\prime}})_{88}\big]^2
+4(M_{\tilde \eta \mathchar`- \tilde \eta^{\prime}})_{08}^2}} \,.
\end{gather}
The eigenstates of the mass matrix (\ref{mass-matrix}) denoted as $\varphi_1$ and $\varphi_2$ are related to $\tilde{\eta}^{\prime}$ and $\tilde{\eta}$ by%
\footnote{In the literature, $\eta'$ and $\eta$ in our notation here are often denoted by
$\eta_0$ and $\eta_8$, and $\varphi_1$ and $\varphi_2$ by $\eta'$ and $\eta$, respectively.}
\begin{gather}
\left(
\begin{array}{c}
\tilde{\eta}^{\prime} \\
\tilde{\eta}
\end{array}
\right)
= \left(
\begin{array}{cc}
{\rm cos}\theta & -{\rm sin}\theta \\
{\rm sin}\theta & {\rm cos}\theta
\end{array}
\right)\, \left(
\begin{array}{c}
\varphi_1 \\
\varphi_2
\end{array}
\right) \,.
\end{gather}
In particular, one can see from eq.~(\ref{mass-eigenvalue}) that the mass eigenvalue for $\varphi_2$ is larger than that for $\varphi_1$.

We now focus on $\varphi_1$. Then, $\Sigma$ can be approximated as
\begin{gather}
\Sigma \approx {\rm diag}\left(
{\rm e}^{{\rm i}\lambda_1\varphi_1} \,, {\rm e}^{{\rm i}\lambda_1\varphi_1} \,, 
{\rm e}^{{\rm i}\lambda_2\varphi_1}
\right) \,, \\
\lambda_1
\equiv \sqrt{\frac{2}{3}} \frac{{\rm cos}\theta}{f_{\tilde{\eta}^{\prime}}} + \frac{1}{\sqrt 3} \frac{{\rm sin}\theta}{f_{\tilde{\pi}}} 
\,, \qquad
\lambda_2 
\equiv \sqrt{\frac{2}{3}} \frac{{\rm cos}\theta}{f_{\tilde{\eta}^{\prime}}} - \frac{2}{\sqrt 3} \frac{{\rm sin}\theta}{f_{\tilde{\pi}}} \notag \,.
\end{gather}
Accordingly, we arrive at the low-energy effective Hamiltonian  
\begin{gather}
\label{mixing Hamiltonian}
\tilde{\mathcal H}_{\rm CFL}
= \frac{1}{2}v^2_{\tilde{\eta}^{\prime}}(\del_z\varphi_1)^2 
+ 4c m_{\rm ud} m_{\rm s}{\rm cos}(\lambda_1\varphi_1) + 2c m_{\rm ud}^2{\rm cos}(\lambda_2\varphi_1)
- \frac{\mu_{\rm B}^2{\rm cos}\theta}
{8\pi^2 f_{\tilde{\eta}^{\prime}}} \sqrt{\frac{2}{3}} \Omega \del_z\varphi_1 \,,
\end{gather}
where $\lambda_{1,2}$ are functions of $m_{\rm ud}$, $m_{\rm s}$, $f_{\tilde{\pi}}$, and $f_{\tilde{\eta}^{\prime}}$. 
The equation of motion for $\varphi_1$ is given by
\begin{gather}
\label{EOM-diag}
v_{\tilde{\eta}^{\prime}}^2\del^2_z\varphi_1
= 4c m_{\rm ud} m_{\rm s} \lambda_1 \, {\rm sin}(\lambda_1\varphi_1)
+ 2c m_{\rm ud}^2\lambda_2 \, {\rm sin}(\lambda_2\varphi_1) \,.
\end{gather}
To the best of our knowledge, this differential equation cannot be analytically solved unlike the flavor symmetric case, and we cannot repeat the previous discussion. In fact, the solution of eq.~(\ref{EOM-diag}) has a periodicity in $\varphi_1$ only if the ratio $\lambda_1/\lambda_2$ is a rational number. As $\lambda_1/\lambda_2$ is generically an irrational number, the periodicity is lost and the solution is no longer CSL in this case.%
\footnote{Note however that this does not mean that the $\eta'$ CSL is absent for {\it any} nonzero $m_{\rm s} - m_{\rm ud}$. When $m_{\rm ud} \simeq m_{\rm s}$, $\tilde \eta$ and $\tilde \eta^{\prime}$ are almost degenerate and integrating out $\varphi_2$ above is not well justified. Physically, one expects that the $\eta'$ CSL persists as long as $m_{\rm s} - m_{\rm ud}$ is so small that it can be treated as a perturbation.}

\bibliographystyle{jhep}
\bibliography{reference.bib}

\providecommand{\href}[2]{#2}\begingroup\raggedright\begin{thebibliography}{10}

\bibitem{STAR:2017ckg}
{\scshape STAR} collaboration, \emph{{Global $\Lambda$ hyperon polarization in
  nuclear collisions: evidence for the most vortical fluid}},
  \href{https://doi.org/10.1038/nature23004}{\emph{Nature} {\bfseries 548}
  (2017) 62} [\href{https://arxiv.org/abs/1701.06657}{{\ttfamily 1701.06657}}].

\bibitem{Chen:2015hfc}
H.-L. Chen, K.~Fukushima, X.-G. Huang and K.~Mameda, \emph{{Analogy between
  rotation and density for Dirac fermions in a magnetic field}},
  \href{https://doi.org/10.1103/PhysRevD.93.104052}{\emph{Phys. Rev.}
  {\bfseries D93} (2016) 104052}
  [\href{https://arxiv.org/abs/1512.08974}{{\ttfamily 1512.08974}}].

\bibitem{Ebihara:2016fwa}
S.~Ebihara, K.~Fukushima and K.~Mameda, \emph{{Boundary effects and gapped
  dispersion in rotating fermionic matter}},
  \href{https://doi.org/10.1016/j.physletb.2016.11.010}{\emph{Phys. Lett.}
  {\bfseries B764} (2017) 94}
  [\href{https://arxiv.org/abs/1608.00336}{{\ttfamily 1608.00336}}].

\bibitem{Jiang:2016wvv}
Y.~Jiang and J.~Liao, \emph{{Pairing Phase Transitions of Matter under
  Rotation}}, \href{https://doi.org/10.1103/PhysRevLett.117.192302}{\emph{Phys.
  Rev. Lett.} {\bfseries 117} (2016) 192302}
  [\href{https://arxiv.org/abs/1606.03808}{{\ttfamily 1606.03808}}].

\bibitem{Chernodub:2016kxh}
M.~N. Chernodub and S.~Gongyo, \emph{{Interacting fermions in rotation: chiral
  symmetry restoration, moment of inertia and thermodynamics}},
  \href{https://doi.org/10.1007/JHEP01(2017)136}{\emph{JHEP} {\bfseries 01}
  (2017) 136} [\href{https://arxiv.org/abs/1611.02598}{{\ttfamily
  1611.02598}}].

\bibitem{Chernodub:2017ref}
M.~N. Chernodub and S.~Gongyo, \emph{{Effects of rotation and boundaries on
  chiral symmetry breaking of relativistic fermions}},
  \href{https://doi.org/10.1103/PhysRevD.95.096006}{\emph{Phys. Rev.}
  {\bfseries D95} (2017) 096006}
  [\href{https://arxiv.org/abs/1702.08266}{{\ttfamily 1702.08266}}].

\bibitem{Liu:2017zhl}
Y.~Liu and I.~Zahed, \emph{{Rotating Dirac fermions in a magnetic field in 1+2
  and 1+3 dimensions}},
  \href{https://doi.org/10.1103/PhysRevD.98.014017}{\emph{Phys.\ Rev.\ D}
  {\bfseries 98} (2018) 014017}
  [\href{https://arxiv.org/abs/1710.02895}{{\ttfamily 1710.02895}}].

\bibitem{Zhang:2018ome}
H.~Zhang, D.~Hou and J.~Liao, \emph{{Mesonic Condensation in Isospin Matter
  under Rotation}},  \href{https://arxiv.org/abs/1812.11787}{{\ttfamily
  1812.11787}}.

\bibitem{Wang:2018zrn}
L.~Wang, Y.~Jiang, L.~He and P.~Zhuang, \emph{{Local suppression and
  enhancement of the pairing condensate under rotation}},
  \href{https://doi.org/10.1103/PhysRevC.100.034902}{\emph{Phys.\ Rev.\ C}
  {\bfseries 100} (2019) 034902}
  [\href{https://arxiv.org/abs/1901.00804}{{\ttfamily 1901.00804}}].

\bibitem{Chen:2019tcp}
H.-L. Chen, X.-G. Huang and K.~Mameda, \emph{{Do charged pions condense in a
  magnetic field with rotation?}},
  \href{https://arxiv.org/abs/1910.02700}{{\ttfamily 1910.02700}}.

\bibitem{Huang:2017pqe}
X.-G. Huang, K.~Nishimura and N.~Yamamoto, \emph{{Anomalous effects of dense
  matter under rotation}},
  \href{https://doi.org/10.1007/JHEP02(2018)069}{\emph{JHEP} {\bfseries 02}
  (2018) 069} [\href{https://arxiv.org/abs/1711.02190}{{\ttfamily
  1711.02190}}].

\bibitem{Dzyaloshinsky:1964dz}
I.~E. Dzyaloshinsky, \emph{{Theory of helicoidal structures in
  antiferromagnets. I. Nonmetals}}, {\emph{Sov. Phys. JETP} {\bfseries 19}
  (1964) 960}.

\bibitem{togawa2012chiral}
Y.~Togawa, T.~Koyama, K.~Takayanagi, S.~Mori, Y.~Kousaka, J.~Akimitsu et~al.,
  \emph{Chiral magnetic soliton lattice on a chiral helimagnet},
  {\emph{Physical review letters} {\bfseries 108} (2012) 107202}.

\bibitem{Gennes}
P.~G. De~Gennes, \emph{{Calcul de la distorsion d'une structure cholesterique
  par un champ magnetique}},
  \href{https://doi.org/10.1016/0038-1098(68)90024-0}{\emph{Solid State
  Commun.} {\bfseries 128} (1968) 163}.

\bibitem{Brauner:2016pko}
T.~Brauner and N.~Yamamoto, \emph{{Chiral Soliton Lattice and Charged Pion
  Condensation in Strong Magnetic Fields}},
  \href{https://doi.org/10.1007/JHEP04(2017)132}{\emph{JHEP} {\bfseries 04}
  (2017) 132} [\href{https://arxiv.org/abs/1609.05213}{{\ttfamily
  1609.05213}}].

\bibitem{Brauner:2019rjg}
T.~Brauner, G.~Filios and H.~Kole$\check{\rm s}$ov\'{a}, \emph{{Anomaly-Induced
  Inhomogeneous Phase in Quark Matter without the Sign Problem}},
  \href{https://doi.org/10.1103/PhysRevLett.123.012001}{\emph{Phys.\ Rev.\
  Lett.} {\bfseries 123} (2019) 012001}
  [\href{https://arxiv.org/abs/1902.07522}{{\ttfamily 1902.07522}}].

\bibitem{Brauner:2019aid}
T.~Brauner, G.~Filios and H.~Kole$\check{\rm s}$ov\'{a}, \emph{{Chiral soliton
  lattice in QCD-like theories}},
  \href{https://doi.org/10.1007/JHEP12(2019)029}{\emph{JHEP} {\bfseries 12}
  (2019) 029} [\href{https://arxiv.org/abs/1905.11409}{{\ttfamily
  1905.11409}}].

\bibitem{kishine2015theory}
J.-i. Kishine and A.~Ovchinnikov, \emph{Theory of monoaxial chiral helimagnet},
   in \emph{Solid State Physics}, vol.~66, pp.~1--130, Elsevier, (2015).

\bibitem{Son:2004tq}
D.~T. Son and A.~R. Zhitnitsky, \emph{{Quantum anomalies in dense matter}},
  \href{https://doi.org/10.1103/PhysRevD.70.074018}{\emph{Phys. Rev.}
  {\bfseries D70} (2004) 074018}
  [\href{https://arxiv.org/abs/hep-ph/0405216}{{\ttfamily hep-ph/0405216}}].

\bibitem{Son:2007ny}
D.~T. Son and M.~A. Stephanov, \emph{{Axial anomaly and magnetism of nuclear
  and quark matter}},
  \href{https://doi.org/10.1103/PhysRevD.77.014021}{\emph{Phys. Rev.}
  {\bfseries D77} (2008) 014021}
  [\href{https://arxiv.org/abs/0710.1084}{{\ttfamily 0710.1084}}].

\bibitem{Schafer:1998ef}
T.~Sch$\ddot{\rm a}$fer and F.~Wilczek, \emph{{Continuity of quark and hadron
  matter}}, \href{https://doi.org/10.1103/PhysRevLett.82.3956}{\emph{Phys.\
  Rev.\ Lett.} {\bfseries 82} (1999) 3956}
  [\href{https://arxiv.org/abs/hep-ph/9811473}{{\ttfamily hep-ph/9811473}}].

\bibitem{Hatsuda:2006ps}
T.~Hatsuda, M.~Tachibana, N.~Yamamoto and G.~Baym, \emph{{New critical point
  induced by the axial anomaly in dense QCD}},
  \href{https://doi.org/10.1103/PhysRevLett.97.122001}{\emph{Phys. Rev. Lett.}
  {\bfseries 97} (2006) 122001}
  [\href{https://arxiv.org/abs/hep-ph/0605018}{{\ttfamily hep-ph/0605018}}].

\bibitem{Yamamoto:2007ah}
N.~Yamamoto, M.~Tachibana, T.~Hatsuda and G.~Baym, \emph{{Phase structure,
  collective modes, and the axial anomaly in dense QCD}},
  \href{https://doi.org/10.1103/PhysRevD.76.074001}{\emph{Phys. Rev.}
  {\bfseries D76} (2007) 074001}
  [\href{https://arxiv.org/abs/0704.2654}{{\ttfamily 0704.2654}}].

\bibitem{Manes:2019fyw}
J.~L. Ma$\tilde{\rm n}$es, E.~Meg\'{i}as, M.~Valle and M.~{\'A}.
  V\'{a}zquez-Mozo, \emph{{Anomalous Currents and Constitutive Relations of a
  Chiral Hadronic Superfluid}},
  \href{https://doi.org/10.1007/JHEP12(2019)018}{\emph{JHEP} {\bfseries 12}
  (2019) 018} [\href{https://arxiv.org/abs/1910.04013}{{\ttfamily
  1910.04013}}].

\bibitem{Vilenkin:1979ui}
A.~Vilenkin, \emph{{Macroscopic parity violating effects: neutrino fluxes from
  rotating black holes and in rotating thermal radiation}},
  \href{https://doi.org/10.1103/PhysRevD.20.1807}{\emph{Phys. Rev.} {\bfseries
  D20} (1979) 1807}.

\bibitem{Vilenkin:1980zv}
A.~Vilenkin, \emph{{Quantum field theory at finite temperature in a rotating
  system}}, \href{https://doi.org/10.1103/PhysRevD.21.2260}{\emph{Phys. Rev.}
  {\bfseries D21} (1980) 2260}.

\bibitem{Son:2009tf}
D.~T. Son and P.~Sur\'{o}wka, \emph{{Hydrodynamics with Triangle Anomalies}},
  \href{https://doi.org/10.1103/PhysRevLett.103.191601}{\emph{Phys. Rev. Lett.}
  {\bfseries 103} (2009) 191601}
  [\href{https://arxiv.org/abs/0906.5044}{{\ttfamily 0906.5044}}].

\bibitem{Landsteiner:2011tf}
K.~Landsteiner, E.~Meg\'{i}as, L.~Melgar and F.~Pena-Benitez,
  \emph{{Gravitational Anomaly and Hydrodynamics}},
  \href{https://doi.org/10.1088/1742-6596/343/1/012073}{\emph{J. Phys. Conf.
  Ser.} {\bfseries 343} (2012) 012073}
  [\href{https://arxiv.org/abs/1111.2823}{{\ttfamily 1111.2823}}].

\bibitem{Landsteiner:2012kd}
K.~Landsteiner, E.~Meg{\'i}as and F.~Pe{\~{n}}a-Benitez, \emph{Anomalous
  Transport from Kubo Formulae}, pp.~433--468.
\newblock Springer Berlin Heidelberg, Berlin, Heidelberg, 2013.

\bibitem{Landsteiner:2016led}
K.~Landsteiner, \emph{{Notes on Anomaly Induced Transport}},
  \href{https://doi.org/10.5506/APhysPolB.47.2617}{\emph{Acta Phys. Polon.}
  {\bfseries B47} (2016) 2617}
  [\href{https://arxiv.org/abs/1610.04413}{{\ttfamily 1610.04413}}].

\bibitem{Wess:1971yu}
J.~Wess and B.~Zumino, \emph{{Consequences of anomalous Ward identities}},
  \href{https://doi.org/10.1016/0370-2693(71)90582-X}{\emph{Phys. Lett.}
  {\bfseries 37B} (1971) 95}.

\bibitem{Witten:1983tw}
E.~Witten, \emph{{Global Aspects of Current Algebra}},
  \href{https://doi.org/10.1016/0550-3213(83)90063-9}{\emph{Nucl. Phys.}
  {\bfseries B223} (1983) 422}.

\bibitem{Vilenkin:1980fu}
A.~Vilenkin, \emph{{Equilibrium parity violating current in a magnetic field}},
  \href{https://doi.org/10.1103/PhysRevD.22.3080}{\emph{Phys.\ Rev.\ D}
  {\bfseries 22} (1980) 3080}.

\bibitem{Nielsen:1983rb}
H.~B. Nielsen and M.~Ninomiya, \emph{{Adler-Bell-Jackiw anomaly and Weyl
  fermions in crystal}},
  \href{https://doi.org/10.1016/0370-2693(83)91529-0}{\emph{Phys.\ Lett.\ B}
  {\bfseries 130} (1983) 389}.

\bibitem{Fukushima:2008xe}
K.~Fukushima, D.~E. Kharzeev and H.~J. Warringa, \emph{{The Chiral Magnetic
  Effect}}, \href{https://doi.org/10.1103/PhysRevD.78.074033}{\emph{Phys.\
  Rev.\ D} {\bfseries 78} (2008) 074033}
  [\href{https://arxiv.org/abs/0808.3382}{{\ttfamily 0808.3382}}].

\bibitem{Aoki:2014moa}
S.~Aoki and M.~Creutz, \emph{{Pion Masses in Two-Flavor QCD with $\eta$
  Condensation}},
  \href{https://doi.org/10.1103/PhysRevLett.112.141603}{\emph{Phys. Rev. Lett.}
  {\bfseries 112} (2014) 141603}
  [\href{https://arxiv.org/abs/1402.1837}{{\ttfamily 1402.1837}}].

\bibitem{Witten:1980sp}
E.~Witten, \emph{{Large N Chiral Dynamics}},
  \href{https://doi.org/10.1016/0003-4916(80)90325-5}{\emph{Annals Phys.}
  {\bfseries 128} (1980) 363}.

\bibitem{Alford:1998mk}
M.~G. Alford, K.~Rajagopal and F.~Wilczek, \emph{{Color flavor locking and
  chiral symmetry breaking in high density QCD}},
  \href{https://doi.org/10.1016/S0550-3213(98)00668-3}{\emph{Nucl. Phys.}
  {\bfseries B537} (1999) 443}
  [\href{https://arxiv.org/abs/hep-ph/9804403}{{\ttfamily hep-ph/9804403}}].

\bibitem{Son:2000fh}
D.~T. Son, M.~A. Stephanov and A.~R. Zhitnitsky, \emph{{Domain walls of high
  density QCD}}, \href{https://doi.org/10.1103/PhysRevLett.86.3955}{\emph{Phys.
  Rev. Lett.} {\bfseries 86} (2001) 3955}
  [\href{https://arxiv.org/abs/hep-ph/0012041}{{\ttfamily hep-ph/0012041}}].

\bibitem{Son:1999cm}
D.~T. Son and M.~A. Stephanov, \emph{{Inverse meson mass ordering in color
  flavor locking phase of high density QCD}},
  \href{https://doi.org/10.1103/PhysRevD.61.074012}{\emph{Phys. Rev.}
  {\bfseries D61} (2000) 074012}
  [\href{https://arxiv.org/abs/hep-ph/9910491}{{\ttfamily hep-ph/9910491}}],
  [{\it Erratum ibid.} {\bf 62} (2000) 059902].

\bibitem{Beane:2000ms}
S.~R. Beane, P.~F. Bedaque and M.~J. Savage, \emph{{Meson masses in high
  density QCD}},
  \href{https://doi.org/10.1016/S0370-2693(00)00606-7}{\emph{Phys. Lett.}
  {\bfseries B483} (2000) 131}
  [\href{https://arxiv.org/abs/hep-ph/0002209}{{\ttfamily hep-ph/0002209}}].

\bibitem{Son:2000tu}
D.~T. Son and M.~A. Stephanov, \emph{{Inverse meson mass ordering in color
  flavor locking phase of high density QCD: Erratum}},
  \href{https://doi.org/10.1103/PhysRevD.62.059902}{\emph{Phys. Rev.}
  {\bfseries D62} (2000) 059902}
  [\href{https://arxiv.org/abs/hep-ph/0004095}{{\ttfamily hep-ph/0004095}}].

\bibitem{Schafer:2002ty}
T.~Sch$\ddot{\rm a}$fer, \emph{{Instanton effects in QCD at high baryon
  density}}, \href{https://doi.org/10.1103/PhysRevD.65.094033}{\emph{Phys.
  Rev.} {\bfseries D65} (2002) 094033}
  [\href{https://arxiv.org/abs/hep-ph/0201189}{{\ttfamily hep-ph/0201189}}].

\bibitem{Yamamoto:2008zw}
N.~Yamamoto, \emph{{Instanton-induced crossover in dense QCD}},
  \href{https://doi.org/10.1088/1126-6708/2008/12/060}{\emph{JHEP} {\bfseries
  12} (2008) 060} [\href{https://arxiv.org/abs/0810.2293}{{\ttfamily
  0810.2293}}].

\bibitem{Donoghue:1992dd}
J.~F. Donoghue, E.~Golowich and B.~R. Holstein, \emph{Dynamics of the Standard
  Model}, Cambridge Monographs on Particle Physics, Nuclear Physics and
  Cosmology. Cambridge University Press, 1992,
  \href{https://doi.org/10.1017/CBO9780511524370}{10.1017/CBO9780511524370}.

\bibitem{Brauner:2017mui}
T.~Brauner and S.~Kadam, \emph{{Anomalous electrodynamics of neutral pion
  matter in strong magnetic fields}},
  \href{https://doi.org/10.1007/JHEP03(2017)015}{\emph{JHEP} {\bfseries 03}
  (2017) 015} [\href{https://arxiv.org/abs/1701.06793}{{\ttfamily
  1701.06793}}].

\bibitem{tHooft:1976snw}
G.~'t~Hooft, \emph{{Computation of the Quantum Effects Due to a
  Four-Dimensional Pseudoparticle}},
  \href{https://doi.org/10.1103/PhysRevD.18.2199.3,
  10.1103/PhysRevD.14.3432}{\emph{Phys. Rev.} {\bfseries D14} (1976) 3432}
  [{\it Erratum ibid.} {\bf 18} (1978) 2199].

\bibitem{Shifman:1979uw}
M.~A. Shifman, A.~I. Vainshtein and V.~I. Zakharov, \emph{{Instanton Density in
  a Theory with Massless Quarks}},
  \href{https://doi.org/10.1016/0550-3213(80)90389-2}{\emph{Nucl. Phys.}
  {\bfseries B163} (1980) 46}.

\bibitem{Schafer:1996wv}
T.~Sch$\ddot{\rm a}$fer and E.~V. Shuryak, \emph{{Instantons in QCD}},
  \href{https://doi.org/10.1103/RevModPhys.70.323}{\emph{Rev. Mod. Phys.}
  {\bfseries 70} (1998) 323}
  [\href{https://arxiv.org/abs/hep-ph/9610451}{{\ttfamily hep-ph/9610451}}].

\bibitem{Shuryak:1982hk}
E.~V. Shuryak, \emph{{The Role of Instantons in Quantum Chromodynamics. 3.
  Quark - Gluon Plasma}},
  \href{https://doi.org/10.1016/0550-3213(82)90480-1}{\emph{Nucl. Phys.}
  {\bfseries B203} (1982) 140}.

\bibitem{Schafer:1999fe}
T.~Sch$\ddot{\rm a}$fer, \emph{{Patterns of symmetry breaking in QCD at high
  baryon density}},
  \href{https://doi.org/10.1016/S0550-3213(00)00063-8}{\emph{Nucl. Phys.}
  {\bfseries B575} (2000) 269}
  [\href{https://arxiv.org/abs/hep-ph/9909574}{{\ttfamily hep-ph/9909574}}].

\bibitem{Manuel:2000wm}
C.~Manuel and M.~H. Tytgat, \emph{{Masses of the Goldstone modes in the CFL
  phase of QCD at finite density}},
  \href{https://doi.org/10.1016/S0370-2693(00)00331-2}{\emph{Phys.\ Lett.\ B}
  {\bfseries 479} (2000) 190}
  [\href{https://arxiv.org/abs/hep-ph/0001095}{{\ttfamily hep-ph/0001095}}].

\end{thebibliography}\endgroup


\end{document}